\documentclass[sigconf]{acmart} 

\usepackage{etoolbox,environ}

\copyrightyear{2022}
\acmYear{2022}
\setcopyright{rightsretained}
\acmConference[ICSE '22]{44th International Conference on Software Engineering}{May 21--29, 2022}{Pittsburgh, PA, USA}
\acmBooktitle{44th International Conference on Software Engineering (ICSE '22), May 21--29, 2022, Pittsburgh, PA, USA}
\acmDOI{10.1145/3510003.3510082}
\acmISBN{978-1-4503-9221-1/22/05}

\usepackage{listings}
\usepackage{xcolor}
\usepackage{xspace}
\usepackage{enumitem}
\usepackage{hyperref}
\usepackage{tikz-uml}
\usepackage{microtype}
\usepackage{tabu}
\usepackage{multirow}
\usepackage{threeparttable}
\usepackage{pifont}
\usepackage[justification=centering]{caption}
\usepackage{setspace}
\usepackage{supertabular,booktabs}
\usepackage{longtable}
\usepackage[linesnumbered,ruled,lined]{algorithm2e}
\usepackage{makecell}
\usepackage{amsmath}
\usepackage{graphicx}
\usepackage{tcolorbox}

\newcommand{\tool}{{\sc LTL-Fuzzer}\xspace}
\newcommand{\toolBold}{{\sc\bfseries LTL-Fuzzer}\xspace}
\newcommand{\aflgo}{{\sc AFLGo}\xspace}
\newcommand{\aflgoBold}{{\sc\bfseries AFLGo}\xspace}
\newcommand{\paflgo}{{\sc $\rm AFL_{LTL}$}\xspace}
\newcommand{\paflgoBold}{{\sc\bfseries $\rm AFL_{LTL}$}\xspace}

\newcommand{\checker}{{\sc L+NuSMV}\xspace}
\newcommand{\checkerBold}{{\sc\bfseries L+NuSMV}\xspace}

\usepackage{tabularx,adjustbox,booktabs}
\usepackage{blindtext}

\newcommand\clearrow{\global\let\rowmac\relax}

\definecolor{mycolor}{rgb}{0.122, 0.435, 0.698}
\newcommand{\result}[1]{%
\begin{tcolorbox}[colframe=mycolor,boxrule=0.5pt,arc=4pt,
      left=6pt,right=6pt,top=6pt,bottom=6pt,boxsep=0pt,width=\columnwidth]%
      {#1}
\end{tcolorbox}%
}

\newcommand{\todo}[1]{}
\renewcommand{\todo}[1]{{\color{red} TODO: {#1}}}

\begin{document}

\title{Linear-time Temporal Logic guided Greybox Fuzzing}

\author{Ruijie Meng}
\authornote{Joint first authors}
\affiliation{%
  \institution{National University of Singapore}
  \country{Singapore}
  }
\email{ruijie@comp.nus.edu.sg}

\author{Zhen Dong}
\authornotemark[1]
\authornote{Corresponding Author}
\authornote{Zhen Dong and Ivan Beschastnikh participated in the work while being at the National University of Singapore as post-doc and visiting Associate Professor respectively.}
\affiliation{%
  \institution{Fudan University}
  \country{China}
  }
\email{zhendong@fudan.edu.cn}

\author{Jialin Li}
\affiliation{%
  \institution{National University of Singapore}
  \country{Singapore}
  }
\email{lijl@comp.nus.edu.sg}

\author{Ivan Beschastnikh}
\authornotemark[3]
\affiliation{%
  \institution{University of British Columbia}
  \country{Canada}
  }
\email{bestchai@cs.ubc.ca}

\author{Abhik Roychoudhury}
\affiliation{%
  \institution{National University of Singapore}
  \country{Singapore}
  }
\email{abhik@comp.nus.edu.sg}

\begin{abstract}
Software model checking as well as runtime verification are verification techniques which are widely used for checking temporal properties of software systems. Even though they are  property verification techniques, their common usage in practice is in "bug finding", that is, finding violations of temporal properties. Motivated by this observation and leveraging the recent progress in fuzzing, we build a greybox fuzzing framework to find violations of Linear-time Temporal Logic (LTL) properties.

Our framework takes as input a sequential program written in C/C++, and an LTL property. It finds violations, or counterexample traces, of the LTL property in stateful software systems; however, it does not achieve verification. Our work substantially extends directed greybox fuzzing to witness arbitrarily complex event orderings. We note that existing directed greybox fuzzing approaches are limited to witnessing reaching a location or witnessing simple event orderings like use-after-free. At the same time, compared to model checkers, our approach finds the counterexamples faster, thereby finding more counterexamples within a given time budget.

Our \tool tool, built on top of the AFL fuzzer, is shown to be effective in detecting bugs in  well-known protocol implementations, such as OpenSSL and Telnet. We use \tool to reproduce known vulnerabilities (CVEs), to find 15 zero-day bugs by checking properties extracted from RFCs (for which 12 CVEs have been assigned), and to find violations of both safety as well as liveness properties in real-world protocol implementations. Our work represents a practical advance over software model checkers --- while simultaneously representing a conceptual advance over existing greybox fuzzers. Our work thus provides a starting point for understanding the unexplored synergies 
among software model checking, runtime verification and greybox fuzzing.
\end{abstract}

\maketitle

\section{Introduction\label{sec:intro}}

Software model checking is a popular validation and verification method for reactive stateful software systems. It is an automated technique to check temporal logic properties (constraining event orderings in program execution) against a finite state transition system. Model checking usually suffers from the state space explosion problem; this is exacerbated in software systems which are naturally infinite-state. To cope with infinitely many states, the research community has looked into automatically deriving a hierarchy of finite state abstractions via predicate abstractions and abstraction refinement of the program's data memory (e.g. see ~\cite{slam}). Whenever a counterexample trace is found in such model checking runs, the trace can be analyzed to find (a) whether it is a spurious counterexample introduced due to abstractions, or (b) the root-cause / bug causing the counterexample. This has rendered model checking to be a useful automated bug finding method for software systems.

Runtime verification is a lightweight and yet rigorous verification method, which complements model checking \cite{introductionRV, briefRV, studyRV}. In runtime verification, a single execution of a system is dynamically checked against formally specified properties (e.g, temporal logic properties). Specifically, formal properties specify the correct behaviours of a system. Then the system is instrumented to capture events that are related to the properties being checked. During runtime, a monitor 
collects the events to generate execution traces and checks whether the traces conform to the specified properties. When the properties are violated, it reports violations. Runtime verification aims to achieve a lightweight but not full-fledged verification method. It verifies software systems at runtime without the need of constructing models about software systems and execution environments. However, to generate effective execution traces, software systems are required to be fed many inputs. These inputs are usually obtained manually or via random generation \cite{studyRV}; therefore, runtime verification may take much manual effort and explore many useless inputs in the process of exposing property violations.

Parallel to the works in software model checking and runtime verification, greybox fuzzing methods~\cite{afl,libfuzzer} have seen substantial recent advances. These methods conduct a biased random search over the domain of program inputs, to find bugs or vulnerabilities. The main advantage of greybox fuzzing lies in its scalability to large software systems. However, greybox fuzzing is only a testing (not verification) method and it is mostly useful for finding witnesses to simple oracles such as crashes or overflows. Recently there have been some extension of greybox fuzzing methods towards generating witnesses of more complex oracles, such as tests reaching a location~\cite{aflgo}. However, generating inputs and traces satisfying a complex temporal property remains beyond the reach of current greybox fuzzing tools. Thus, today's greybox fuzzing technology cannot replace the bug-finding abilities of software model checking and runtime verification.

In this paper, we take a step forward in understanding the synergies among software model checking, runtime verification and greybox fuzzing. Given a sequential program and a Linear-time Temporal Logic (LTL) property $\phi$, we construct the B{\"u}chi automata ${\mathcal{A}}_{\neg\phi}$
accepting $\neg \phi$, and use this automata to guide the fuzz campaign. Thus, given a random input exercising an execution trace $\pi$, we can check the "progress" of $\pi$ in reaching the accepting states of ${\mathcal{A}}_{\neg\phi}$, and derive from ${\mathcal{A}}_{\neg\phi}$, the  events that are needed to make further progress in the automata. Furthermore, in general, traces accepted by ${\mathcal{A}}_{\neg\phi}$ are infinite in length and visit an accepting state infinitely often. To accomplish the generation of such infinite-length traces in the course of a fuzz campaign, we can take application state snapshots (at selected program locations) and detect whether an accepting state of ${\mathcal{A}}_{\neg\phi}$ is being visited with the same program state. The application state snapshot can also involve a state abstraction if needed, in which case the  counterexample trace can be subsequently validated via concrete execution.

We present a fuzzing-based technique that directs fuzzing to find violations of {\em arbitrary LTL properties}. To the best of our knowledge, no existing fuzzing technique is capable of finding violations of complex constraints on event orderings such as LTL properties. Existing works on greybox fuzzing are limited to finding witnesses of simple properties such as crashes or use-after-free.  This is the main contribution of our work: algorithms and an implementation of our ideas in a tool that is able to validate {\em any LTL property}, thereby covering a much more expressive class of properties than crashes or use-after-free. Our work adapts directed greybox fuzzing (which directs the search towards specific program locations) to find violations of temporal logic formulae.
We realize our approach for detecting violations of LTL properties in a new greybox fuzzer tool called \tool. \tool is built on top of the AFL fuzzer~\cite{afl} and involves additional program instrumentation to check if a particular execution trace is accepted by the B{\"u}chi automaton representing the negation of the given LTL property.

We evaluated \tool on well-known and large-scale protocol implementations such as OpenSSL, OpenSSH, and Telnet. We show that it efficiently finds bugs that are violations of both safety and liveness properties. We use \tool to reproduce known bugs/violations in the protocol implementations. More importantly, for 50 LTL properties that we manually extracted from Request-for-Comments (RFCs), \tool found 15 new bugs (representing the violation of these properties), out of which 12 CVEs have been assigned. These are zero-day bugs which have previously not been found. We make the data-set of properties and the bugs found available with this paper. We expect that in future, other researchers will take forward the direction in this paper to detect temporal property violations via greybox fuzzing. The data-set of bugs found by \tool can thus form a baseline standard for future research efforts. The dataset and tool are available at {\color{blue}\bf\url{ https://github.com/ltlfuzzer/LTL-Fuzzer}}

\section{Approach overview~\label{sec:approach}}


At a high level, our approach takes a sequential program $P$ and a Linear-time Temporal Logic (LTL) property $\phi$ as inputs. The atomic propositions in $\phi$ refer to predicates over the program variables that can be evaluated to true or false. An example is a predicate $x > y$ in which $x$ and $y$ are program variables. Our approach identifies program locations at which the atomic propositions in the LTL property may be affected. For this, we find program locations at which the values of variables in the atomic proposition and their aliases may change.\footnote{In general, our approach requires an alias analysis to map the atomic propositions to program locations.}
Our technique outputs a {\em counterexample}, i.e., a concrete program input that leads to a violation of the specification. Counterexample generation proceeds in two phases. In the first phase, the program $P$ is transformed into $P^\prime$. 
For this, we use code instrumentation to monitor program behaviors and state transitions during program execution. We check these against the provided LTL property. In the second phase, a fuzz campaign is launched for the program $P^\prime$ to find a counterexample through directed fuzzing.    
%
    %

We illustrate our technique with an FTP implementation called Pure-FTPd\footnote{\url{https://www.pureftpd.org/project/pure-ftpd/}}.
Pure-FTPd is a widely-used open source FTP server which complies with the FTP RFC\footnote{\url{https://www.w3.org/Protocols/rfc959/}}. Here is a property described in the RFC that an FTP implementation must satisfy. The FTP server must stop receiving data from a client and reply with code 552 when user quota is exceeded while receiving data.  Code 552 indicates the allocated storage is exceeded.
Throughout this paper, we will use this FTP property -- as represented by $\phi$ -- to illustrate how our technique finds property violations in Pure-FTPd.

\subsection{LTL Property Construction}

We start by manually translating the informal property in the RFC into a LTL property $\phi$.
For this, we
search the Pure-FTPd source code using keywords \texttt{APPE} and \texttt{552}. 
Source code analysis reveals that (1) Pure-FTPd implements a quota-based mechanism to manage user storage space and it works only when activated, and (2) the command \texttt{APPE} is handled by the function \texttt{dostor()}, in which \texttt{user\_quota\_size} is checked when receiving data. When the quota is exceeded, the server replies with code 552 (\texttt{MSG\_QUOTA\_EXCEEDED}) via the function \texttt{addreply()}. 
We therefore construct the property $\phi$ as 
\begin{equation}
\label{eq:ltl}
\neg F (a \wedge F(o \wedge G \neg n))
\end{equation}
The negation of $\phi$ is thus 
\[F (a \wedge F(o \wedge G\neg n))\] where definition of atomic propositions $a,o,n$ appear in
Table~\ref{tab:map}.

\begin{table}[t]
    \centering
    \footnotesize
    \caption{Mapping between atomic propositions and program locations (``...'' indicates omitted loop entries).}
    \begin{tabular}{|l|c|c|c|}\hline
          Predicate   & Atomic Prop. & Locations  \\ \hline
         $quota\_activated = true$ & $a$ & $\langle ftpd.c,6072\rangle$ \\ \hline 
         $user\_dir\_size > user\_quota$ & $o$ &   \makecell{$\langle safe\_rw.c,12 \rangle$ \\$\langle safe\_rw.c,43 \rangle$}\\ \hline
         $msg\_quota\_exceeded= true$ & $n$ &  \makecell{$\langle ftpd.c, 4444 \rangle$ \\ $\langle ftpd.c, 3481 \rangle$} \\ \hline
         $loop\_entry = true$ & $l$ & $\langle ftpd.c, 4067 \rangle$ ... \\ \hline
    \end{tabular}
    \label{tab:map}
\end{table}

Next, we identify program locations where the values of variables in atomic propositions in $\phi$ may  change at runtime.
A simple example is the proposition $quota\_activated = true$, which
corresponds to the program location where quota checking is enabled in Pure-FTPd. 
At another statement, $user\_dir\_size > user\_quota$, we consider the first statement of functions 
%
%
that are used to store data in user directories.  As a result, whenever data is written to user directories, those functions will be invoked and this proposition will be evaluated, i.e., all cases where user quota is exceeded will be captured in an execution.
%
%
For  $msg\_quota\_exceeded= true$, we identify function invocations of \texttt{addreply(552, MSG\_QUOTA\_EXCEEDED...)} which are a reply to clients when the quota is exceeded.
%
%
Specific program locations for each atomic proposition are listed in Table~\ref{tab:map}. 
%
%
%
Their corresponding code snippets are shown in Listings~\ref{lst:enable},~\ref{lst:overflow},~\ref{lst:notify}, and~\ref{lst:loop}.
Here, we show one code snippet per atomic proposition. 
For convenience, we use a tuple $\langle l, p, c_p\rangle$ in which $l$ denotes a program location, $p$ is an atomic proposition, and $c_p$ represents the predicate for  the atomic proposition $p$.
At the end of our manual LTL property generation process, we output a list $L$ comprising such tuples. 
For the example property, the manual process of writing down the predicates and the accompanying tuples was completed by one of the authors in 20 minutes.

\begin{lstlisting}[caption={Enabling the user quota option:<ftpd.c, 6072>.}, label={lst:enable}]
6063  #ifdef QUOTAS
6064  case 'n': {
...
6072    user_quota_size *= (1024ULL * 1024ULL);
6073 +  if(1){
6074 +    generate_event("a");
6075 +    if(liveness) record_state();
6076 +  }
\end{lstlisting}

\begin{lstlisting}[caption={Writing to user directories:<safe\_rw.c, 12>.}, label={lst:overflow}]
12  safe_write(const int fd, const void * const buf_, 
13      size_t count, const int timeout)
14  {
15 +  if(user_dir_size > user_quota){
16 +    generate_event("o");
17 +    if(liveness) record_state();
18 +  }
 
\end{lstlisting}

\begin{lstlisting}[caption={Replying msg\_quota\_exceeded:<ftpd.c, 4444>.}, label={lst:notify}]
4442  afterquota:
4443    if (overflow > 0) {
4444      addreply(552, MSG_QUOTA_EXCEEDED, name);
4445 +    if(1){
4446 +      generate_event("n");
4447 +      if(liveness) record_state();
4448 +  }
 
\end{lstlisting}

\begin{lstlisting}[caption={Entry of a loop statement:<ftpd.c. 4067>.}, label={lst:loop}]
4066  for (;;) {
4067 +  if(1){
4068 +      generate_event("l");
4069 +      if(liveness) record_state();
4070 +  }
\end{lstlisting}

\subsection{Program Transformation}
\label{sec:transformation}

After deriving the property $\phi$ and the list of tuples $L$, we transform program $P$ into $P^\prime$, which can report a failure at runtime whenever $\phi$ is violated.
We perform this program transformation using two instrumentation modules: (1) {\em Event generator,} which generates an event when  a proposition in $\phi$ is evaluated to true
at runtime; (2) {\em Monitor,} which collects the generated events into an execution trace and evaluates if the trace violates $\phi$. 
If a violation is found, the monitor reports a failure.

\paragraph{Event Generator.} To detect changes in $\phi$'s proposition values during program execution, the event generator injects event generation statements at specific program locations. 
To do so, the generator takes the list $L$ produced in the previous step as input. 
For each tuple $\langle l, p, c_p\rangle \in L$, the generator injects a statement \texttt{if($c_p$) generate\_event("$p$");} at the program location $l$, such that an event associated with $p$ can be generated when condition $c_p$ is satisfied.  
For instance, the program location $\langle ftpd.c, 6072\rangle$ corresponds to the proposition variable $a$ ($quota\_activated = true$) and the enabling condition is \texttt{true}. 
The generator then inserts a statement \texttt{if(1) generate\_event("a");} at line 6072 in ftpd.c (see Listing~\ref{lst:enable}).
Consequently, whenever $\langle ftpd.c, 6072\rangle$ is reached, an event associated with $a$ is generated and recorded at runtime. 
Instrumentation for the other tuples appear in Listings~\ref{lst:overflow},~\ref{lst:notify}, and~\ref{lst:loop}. 

\begin{figure}[!t]
    \centering
    \includegraphics[page=1, trim=0 0.1in 0in 0, clip, scale=.6]{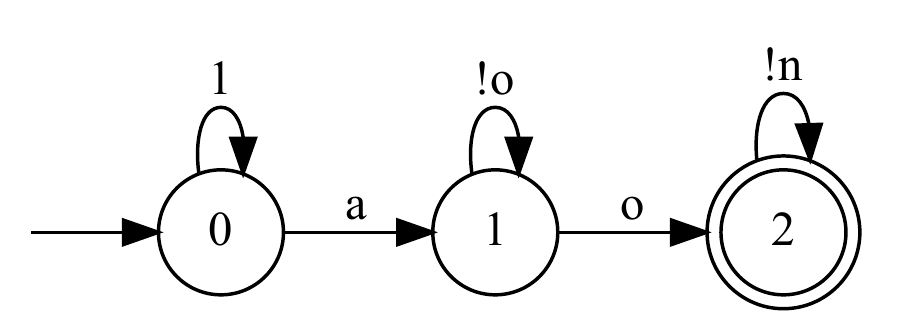}
    \caption{Büchi automata accepting traces satisfying $^\neg \phi$.}
    \label{fig:automata}
    \vspace{-0.3cm}
\end{figure}

\paragraph{Monitor.} The monitor module inserts a \emph{monitor} into program $P$ to verify if the program behavior conforms to property $\phi$ at runtime. 
Specifically, the monitor produces a trace $\tau$ by collecting events that are generated during execution (by the instrumented code). 
It then converts the negation of $\phi$ to a B\"{u}chi automata $\mathcal{A}_{^\neg \phi}$, and checks whether $\mathcal{A}_{^\neg \phi}$ accepts $\tau$. 
If the trace is accepted, the monitor reports a failure, i.e., $\phi$ does not hold in $P$.
In our Pure-FTPd example, the negation of $\phi$ is $F (a \wedge F(o \wedge G \neg n))$, 
%
and the converted B\"{u}chi automata $\mathcal{A}_{^\neg \phi}$ is illustrated in Figure~\ref{fig:automata}. 

\paragraph{Checking Safety Properties.} A B\"{u}chi automata accepts a trace $\tau$ if and only if $\tau$ visits an accepting state of the automata ``infinitely often'' (e.g., state 2 in Figure~\ref{fig:automata}).
For the negation of a \textit{safety} property ($^\neg \phi$), the  B\"{u}chi automata ${\mathcal A}_{\neg\phi}$ accepts all traces which reach an accepting state, since all traces reaching an accepting state will loop there infinitely often. Since only a finite prefix of the trace is relevant for obtaining the counter-example of a safety property, the monitor thus outputs a counterexample if it witnesses a trace that leads to an accepting state in the B{\"u}chi automata ${\mathcal A}_{\neg\phi}$.


\paragraph{Checking Liveness Properties.} 
The B{\"u}chi automata of the negation of $\phi$ accepts a trace $\tau$ if and only if $\tau$ visits an accepting state of $\mathcal{A}_{^\neg \phi}$  ``infinitely often'' (e.g., state 2 in Figure~\ref{fig:automata}). For instance, an infinite trace $a,o,  (v)^\omega$ in which $v \neq n$ will be accepted by $\mathcal{A}_{^\neg \phi}$.  Formally, such a trace has the form $\tau = \tau_1 (\tau_2)^\omega$ ($\left| \tau_2 \right| \neq 0$), where  $\tau_1$ starts in an initial state of the B{\"u}chi automata $\mathcal{A}_{^\neg \phi}$ and runs until an accepting state $s$ of $\mathcal{A}_{^\neg \phi}$, and $\tau_2$ runs from the accepting state $s$ back to itself.  
Witnessing a trace $\tau = \tau_1 (\tau_2)^\omega$ in which $\tau_2$ occurs ``infinitely many times'' is difficult in practice, since a fuzz campaign visits program executions which are necessarily of finite length. A straightforward approach to tackle this difficulty is to detect a loop in the trace and terminate execution when witnessing the loop occurs $m$ times, e.g., $\tau =\tau_1, \overbrace{\tau_2 \tau_2 \cdots}^{m}$. This approach is insufficient because witnessing $\tau_2$ for $m$ times does not guarantee $\tau_2$ occurs infinitely often, for instance \texttt{for (i=0; i<m+2; i++)\{... $\tau_2$...\}} may generate $\tau_2$ for $m$ times but stops generating $\tau_2$ after \texttt{i==m+1}. 

In this paper, we record program states when events associated with atomic propositions occur in the execution and detect a state loop in the witnessed trace. If the execution of the state loop produces $\tau_2$, that means, trace $\tau_2$ can be generated infinitely many times by repeatedly going through the state loop. As a result, we assume that  the witnessed trace can be extended to an infinite $\tau_1 (\tau_2)^\omega$ shaped trace.
Consider following two sequences witnessed in the execution 
\begin{equation*}
\begin{aligned}
 &\tau_e =e_0 e_1 \cdots e_i e_{i+1} \cdots e_{i+h} e_i \cdots e_{i+h} \\
 &\tau_s =s_0 s_1 \cdots \overbrace{s_i s_{i+1} \cdots s_{i+h}}^{loop\, body} s_{i+h+1}  \cdots s_{i+2h}
\end{aligned}
\end{equation*}
where $\tau_e$ is a sequence of events associated with  atomic propositions that occur in the execution and $\tau_s$ is a sequence of program states that are recorded when events occur, for instance $s_i$ indicates the program state that is recorded when the event $e_i$ occurs. Suppose $s_i$ is identical to $s_{i+h+1}$, then $s_i \cdots s_{i+h+1}$ is a state loop and its loop body is $s_i \cdots s_{i+h}$. Whenever $s_i$ takes input $I_{s_i \cdots s_{i+h+1}}$ that leads to $s_i$ from $s_{i+h+1}$, $s_i$ will transition to $s_i$ itself. We assume that the system under test is a reactive system taking a sequence of inputs and it is deterministic, that is, the same input always leads to the same program behavior in the execution.  Thus, $e_i e_{i+1} \cdots e_{i+h}$  can be generated infinitely many times by repeatedly executing input $I_{s_i \cdots s_{i+h+1}}$ on state $s_i$. Trace $\tau_e=e_0\cdots e_{i-1}(e_i\cdots e_{i+h})^\omega$ can be generated by running input $I_{s_0\cdots s_{i}} (I_{s_i \cdots s_{i+h+1}})^\omega$, where $I_{s_0\cdots s_{i}}$ is an input that leads to state $s_{i}$ from $s_0$ and $I_{s_i \cdots s_{i+h+1}}$ is an input that leads to $s_i$ from $s_{i+h+1}$. 

As explained, occurrence of a state loop in the execution is evidence that the witnessed trace can be extended to an infinite $\tau_1 (\tau_2)^\omega$ shaped trace. We leverage this idea to find a violation of a liveness property. When witnessing a trace in the execution that can be extended to a $\tau_1 (\tau_2)^\omega$ shaped trace that is accepted by B{\"u}chi automata $\mathcal{A}_{^\neg \phi}$, we consider a violation of the liveness property has been found. Hence, for liveness property guided fuzzing, we enrich the program transformation of $P$ to $P'$ as follows: (1) instrumenting a function call that records the current program state when an event appearing in a transtion label of 
${\mathcal A}_{\neg\phi}$
occurs in the execution (shown in Listing~\ref{lst:enable}-\ref{lst:loop}) --- specifically, function call \texttt{record\_state()} takes the current program state and generates a hash code for the state at runtime; (2) instrumenting event-generating and state-recording statements at the entries of \texttt{for} and \texttt{while} loop statements in the program to observe possible loops in fuzzing. Listing ~\ref{lst:loop} shows the instrumentation of a \texttt{for} loop statement in Pure-FTPd. More detailed and specific optimizations about state saving for checking liveness properties, appear in
Section \ref{sec:statesaving}.

\subsection{Witnessing Event Sequences}

Since program $P^\prime$, generated in the previous step, reports a failure when $\phi$ is violated, we can find a counterexample for $\phi$ by fuzzing $P^\prime$. An input that leads to such a failure is a counterexample. However, finding an input of this kind is challenging because it has to generate an execution in which certain events occur in a specific order. In our running example of Pure-FTPd, the quota mechanism must be activated first in the execution, then $user\_quota$ must be exceeded, and finally the execution must enter a loop in which no $msg\_quota\_exceeded$ is sent back to the client. Existing directed fuzzing approaches like AFLGo \cite{aflgo} aim to direct fuzzing towards a particular program location and cannot drive execution through multiple program locations in a specific order. We now discuss our B{\"u}chi automata guided fuzzing in the next section.

\section{B{\"u}chi Automata guided fuzzing}
\label{sec:guided-fuzzing}

Given an LTL property $\phi$ to be checked, automata-theoretic model checking of LTL properties \cite{Vardi86} constructs the B{\"u}chi automata ${\mathcal{A}}_{\neg\phi}$ accepting all traces satisfying $\neg\phi$. In this section we will discuss how ${\mathcal{A}}_{\neg\phi}$ can be used to guide fuzzing.
First we design a mechanism to generate an input whose execution passes through multiple program locations in a specific order. 
We design this mechanism by augmenting a greybox fuzzer in two ways.
\begin{itemize}
    \item Power scheduling. During fuzzing, the power scheduling component tends to select seeds {\em closer} to the target on the pre-built inter-procedural control flow graph. Thus, the target can be reached efficiently. To achieve this, we use the fuzzing algorithm of  AFLGo~\cite{aflgo}. 
    \item Input prefix saving. This component observes execution and records input elements that have been consumed when reaching a target.
\end{itemize} 
As mentioned, we focus on fuzzing reactive systems that take a sequence of inputs. 
The mechanism we follow involves directing fuzzing towards multiple program locations in a specific order. 
Consider a sequence of program locations $l_1, l_2\cdots l_m$. Our approach works as follows: first, it takes $l_1$ as the first target and focuses on generating an input that leads to $l_1$. Meanwhile, it observes execution and records the prefix $i_1$ that leads to $l_1$. Next, it takes $l_2$ as the target, and focuses on exploring the space of inputs starting with prefix $i_1$, i.e., keeping generating inputs starting with $i_1$. As a result,  an input that reaches $l_2$ via $l_1$ can be generated.  

Based on the above mechanism of visiting a sequence of program locations, we develop an automata-guided fuzzing approach. The approach uses the B{\"u}chi automata $\mathcal{A}_{^\neg \phi}$ instrumented in program $P^\prime$ and observes the progress that each trace makes on $\mathcal{A}_{^\neg \phi}$ at runtime, e.g., how many state transitions are made towards the accepting state.  To guide fuzzing, the approach saves the progress each input achieves on $\mathcal{A}_{^\neg \phi}$ and uses it to generate inputs that make further progress. Specifically, it saves the progress for each input by recording state transitions that are executed on $\mathcal{A}_{^\neg \phi}$ and the input prefix that leads to those transitions.  Consider input $i_0$ and its trace $\tau_0$ goes from initial state $s_0$ to state $s_m$ on automata $\mathcal{A}_{^\neg \phi}$. The achieved progress is represented as a tuple $\langle x^i_0, x^s_0 \rangle$, where $x^i_0$ is the shortest prefix of $i_0$ whose execution trace goes from $s_0$ to $s_m$ and $x^s_0$ is the state transition sequence $s_0 \cdots s_m$ visited. Such {\em progress} tuples are stored into a set $\mathcal{X}$ and are used to guide fuzzing. 

For input generation, the approach takes a tuple from $\mathcal{X}$ and uses it to generate inputs that makes further progress. Consider a tuple $\langle x^i, x^s\rangle$: $x^s$ records state transitions on automata $\mathcal{A}_{^\neg \phi}$ which input prefix $x^i$ has led to. Thus, we can query $\mathcal{A}_{^\neg \phi}$ with $x^s$ to find a transition that makes further progress, i.e., a state transition that gets closer to an accepting state of $\mathcal{A}_{^\neg \phi}$. In the example, assuming $x^s$ is state 0 in Figure~\ref{fig:automata}, then the transition from state 0 to state 1 will be identified since state 1 is closer to the accepting state 2. Suppose $t$ is the next progressive state transition of $x^s$, then we can further query $\mathcal{A}_{^\neg \phi}$ to obtain atomic propositions that trigger transition $t$. Then, by querying the map between atomic propositions and program locations, we can identify program locations for those atomic propositions. In the example, atomic proposition $a$ triggers transition from state 0 to state 1 and its corresponding program location is $\langle ftpd.c, 6072 \rangle$, as shown in Table~\ref{tab:map}.

From the above we can define criteria for an input to make further progress: (1) its execution has to follow the path that an input prefix $x^i$ has gone through such that the generated trace can go through state transitions $x^s$; and, (2) subsequently the execution reaches one of program locations that are identified above to ensure the generated trace takes a step further in $\mathcal{A}_{^\neg \phi}$.

To generate inputs of this kind, our mechanism for generating inputs that traverse a sequence of program locations in a specific order comes into play. Assume $l_i$ is one of program locations identified above, for making further "progress" in ${\mathcal A}_{\neg \phi}$.
The mechanism takes $l_i$ as the target and keeps generating inputs that start with prefix $x^i$ until generating an input that starts with prefix $x^i$ and subsequently visits location $l_i$. This is how our approach uses tuples in $\mathcal{X}$ to generate inputs that make further "progress" towards an accepting state in the B{\"u}chi automata ${\mathcal A}_{\neg \phi}$. The detailed fuzzing algorithm is now presented.
\section{Fuzzing Algorithm~\label{sec:alg}}

\begin{algorithm}[t]
  \small
  \caption{Counterexample-Guided Fuzzing \label{alg:fuzzing}}

  \SetKwFunction{run}{Fuzz}
  \SetKwFunction{getInitState}{getInitState}
  \SetKwFunction{selectPrefix}{selectPrefix}
  \SetKwFunction{selectAP}{selectTargetAtomicProposition}
  \SetKwFunction{selectLocation}{selectProgramLocationTarget}
  \SetKwFunction{runPrefix}{runPrefix}
  \SetKwFunction{distanceGuidedFuzz}{distanceGuidedFuzz}
  \SetKwFunction{generateInput}{generateInput}
  \SetKwFunction{replacePrefix}{replacePrefix}
  \SetKwFunction{evaluate}{evaluate}
  \SetKwFunction{monitor}{attachHandler}
  \SetKwFunction{runtest}{launchTest}
  \SetKwFunction{getLineNum}{getLineNum}
  \SetKwFunction{getEventQ}{getEventQ}
  \SetKwFunction{getTestingTh}{getTestingThread}
  \SetKwFunction{suspend}{suspend}
  \SetKwFunction{hookAsyncTh}{hookAsyncThread}
  \SetKwFunction{runStatement}{runStatement}
  \SetKwFunction{isWaiting}{isWaiting}
  \SetKwFunction{isDone}{isAllEventHooked}
  \SetKwFunction{getTriggeredEvent}{getEvent}
  \SetKwFunction{isAsync}{isAsync}	
  \SetKwFunction{update}{updateEvent}
  \SetKwFunction{setUpperBound}{setUpperBound}
  \SetKwProg{proc}{Procedure}{}{}
  
  \KwIn{$P^\prime$: The transformation of program under test}
  \KwIn{$\mathcal{A}_{^\neg \phi}$: Automata of negation of property under test }
  \KwIn{$map$: Map between propositions and program locations}
  \KwIn{$flag$: True for liveness properties}
  \KwIn{$total\_time$: Time budget for fuzzing}
  \KwIn{$target\_time$: Time budget for reaching a program location} 
  \vspace{5pt} 
  \proc{\run{$P^\prime$, $\mathcal{A}_{^\neg \phi}$, $map$, $flag$, $total\_time$, $target\_time$}} {
 
    $s_0 \leftarrow \getInitState(\mathcal{A}_{^\neg \phi})$ \; 
    $\mathcal{X} \leftarrow \{\langle \emptyset, s_0\rangle \}$ \tcp*[r]{ Starting with init state of $\mathcal{A}_{^\neg \phi}$}
 
  \For{ $time < total\_time$}{
        $\langle x^i_t, x^s_t\rangle  \leftarrow \selectPrefix(\mathcal{X})$  \;
        $p \leftarrow \selectAP(\mathcal{A}_{^\neg \phi}, x^s_t)$ \;
        $l \leftarrow \selectLocation(map, p)$ \;
        
        \For{$time^\prime < target\_time$}{
            \tcp{$\mathcal{D}$: Feedback of CFG distance}
            \tcp{$S_{power}$: Power schedule algorithm}
            \vspace{5pt}
            $I \leftarrow \generateInput(\mathcal{D}, S_{power})$ \;
            $I^\prime \leftarrow \replacePrefix(I, x^i_t)$ \;
            $d, \langle x^i, x^s\rangle \leftarrow \evaluate(P^\prime, I^\prime, flag)$\;
            $\mathcal{D} \leftarrow \mathcal{D} \cup \{d\}$ \;
            $\mathcal{X} \leftarrow \mathcal{X} \cup \{\langle x^i, x^s\rangle\}$
            
        }
        
	 }
}
\end{algorithm}

Algorithm~\ref{alg:fuzzing} shows the workflow of our counterexample guided fuzzing. To find a counterexample, the algorithm guides fuzzing in two dimensions. First, it prioritizes the exploration of inputs whose execution traces are more likely to be accepted by $\mathcal{A}_{^\neg \phi}$. Specifically, if the  trace of the prefix of an input reaches a state that is closer to an accepting state on $\mathcal{A}_{^\neg \phi}$, then its trace is more likely to be accepted. The algorithm selects input prefixes whose traces have been witnessed to get close to an accepting state and keeps generating inputs starting with them (shown in line~5 and line 10). Secondly, the algorithm focuses on generating inputs whose execution makes further progress on $\mathcal{A}_{^\neg \phi}$. Given an input prefix, the algorithm finds a state transition $t$ that helps us get closer to an accepting state in $\mathcal{A}_{^\neg \phi}$, and finds the atomic propositions  which enable $t$ to be taken (line 6).  For the atomic propositions enabling transition $t$, we identify the corresponding program locations (line 7). Then we attempt to generate inputs that reach the program location in the execution and trigger  the program behavior associated with the atomic proposition. As a result, the generated trace can make further progress in $\mathcal{A}_{^\neg \phi}$. To generate inputs that reach a particular program location, we leverage the algorithm proposed in AFLGo (line 8-14). Its idea is to assign more power to seeds that are {\em closer} to the target on a pre-built control flow graph such that  the generated inputs are more likely to reach the target. The time budget for reaching a target is configurable, via parameter $target\_time$.

For prefix selection (line 5), the algorithm defines a fitness function to compute a fitness value for each prefix tuple. Given a tuple $\langle x^i_t, x^s_t\rangle$, its fitness value is $$ f_t = \frac{l_s}{l_s+l_a} + \frac{1}{l_i}$$ where $l_s$ is the length of $x^s_t$ and $l_a$ is the length of the shortest path from the last state of $x^s_t$ to an accepting state on $\mathcal{A}_{^\neg \phi}$ and $l_i$ is the length of input prefix $x^i_t$. As shown in the formula, a prefix tuple has a higher fitness value if the last state of $x^s_t$ is closer to an accepting state on $\mathcal{A}_{^\neg \phi}$ and the input prefix is shorter. Heuristically, by extending such a prefix, our fuzzing algorithm is more likely to generate an input whose execution trace is accepted by $\mathcal{A}_{^\neg \phi}$.  
Prefix tuples with higher fitness values are prioritized for selection.

For atomic proposition selection (line 6), we adopt a random selection strategy. Consider tuple $\langle x^i_t, x^s_t\rangle$ and the last state of $x^s_t$ is $s_t$, the algorithm identifies atomic propositions that make a progressive transition from $s_t$ on $\mathcal{A}_{^\neg \phi}$ as follows: if state $s_t$ is not an accepting state of $\mathcal{A}_{^\neg \phi}$, any atomic proposition that triggers a transition from $s_t$ towards an accepting state is selected. If state $s_t$ is an accepting state, any atomic proposition that triggers a transition from $s_t$ back to itself is selected. For simplicity, the algorithm randomly selects one from the identified atomic propositions. 
When the selected proposition $p$ has multiple associated program locations, we randomly select one of them as a target. The main consideration for adopting a random strategy is to keep our technique as simple as possible. Moreover, these strategies can be configured in our tool.

\section{State Saving~\label{sec:state}}
\label{sec:statesaving}

In liveness property verification, \tool detects a state loop in the witnessed trace. If a state loop is detected,  \tool assumes the current trace can be extended to a lasso-shaped trace $\tau_1(\tau_2)^\omega$. This works with a {\em concrete}  representation of program states, however in reality state representation of software implementations are always abstracted.
State representations that are too abstract may miss capturing variable states that are relevant to the loop, which leads to false positives. State representations that are too concrete may contain variable states that are irrelevant to the loop such as  a variable for system-clock, which leads to false negatives.
To be practical, \tool takes a snapshot of application's registers and {\em addressable} memory and hashes it into a 32-bit integer, which is recorded as a state. Addressable memory indicates two kinds of objects: (1) global variables (2) objects that are explicitly allocated with functions \texttt{malloc()} and \texttt{alloca()}. Such a convention was also adopted in previous works on infinite loop detection~\cite{loop:mit,loop:cav}. 

Furthermore, \tool only records a program state for selected program locations, not for all program locations. Specifically, we only save states for the program locations associated with the transition labels of the automata ${\mathcal{A}}_{\neg\phi}$ where $\phi$ is the liveness property being checked. 
Note that a  transition label in ${\mathcal{A}}_{\neg\phi}$ is a subset of atomic propositions \cite{VW:94,Vardi86}. The full set of atomic propositions is constructed by taking the atomic propositions appearing in $\phi$ and  embellishing this set with atomic propositions that we introduce for occurrence of each program loop header (such as $l$ in Table~\ref{tab:map}). If the transition label involves a set $L$ of atomic propositions, we track states for only those atomic propositions in $L$ which correspond to loop header occurrences. The goal here is to quickly find possible infinite loops by looking for a loop header being visited with the same program state. Hence for the transition label $!n$ in our running example, we only store states for the program locations corresponding atomic proposition $l$ in Table~\ref{tab:map}.


\begin{lstlisting}[caption={Quota checking:<ftpd.c. 4315>.}, label={lst:bug}]
4315  if(...(max_filesize >= (off_t) 0 && 
            (max_filesize=user_quota_size - quota.size)
            < (off_t) 0 )){
         ...
4322     goto afterquota;
4323  }
\end{lstlisting}

In the example shown in Section~\ref{sec:approach}, \tool  witnesses a state generated at program location $\langle ftpd.c, 4067 \rangle$ (shown in Listing~\ref{lst:loop}) that has been observed before and at the same time the witnessed trace is accepted by $\mathcal{A}_{^\neg \phi}$. In this case, \tool reports a violation of the LTL property $\phi$ shown in Page \pageref{eq:ltl}. To validate if the violation is spurious, we check if the observed state loop can be repeated in the execution. 
Our analysis shows a chunk of data was read during the execution of the state loop and the chunk of data was from a file uploaded by the client. We duplicated the chunk of data in the uploaded file and reran the experiment and found the state loop was repeated. That means the witnessed trace can be extended to a  $\tau_1(\tau_2)^\omega$ shaped trace, which visits the accepting state of the automata accepting $\neg\phi$ (shown in Figure \ref{fig:automata}) infinitely many times. Thus, the reported violation is not spurious.  

We further analyzed the root cause of the violation. It shows there was a logical bug in the quota checking module. As shown in Listing~\ref{lst:bug}, the assignment of \texttt{max\_filesize} occurs in a conditional statement and is never executed due to that \texttt{max\_filesize}'s initial value is  -1. To fix the bug, we created a patch and submitted a pull request on the Github repo of Pure-FTPd, which has been confirmed and verified.

\section{LTL-Fuzzer Implementation \label{sec:implement}}

We implement \tool as an open source tool built on top of AFL, which comprises two main components: instrumentor and fuzzer. In the following, we explain these components.

\subsection{Instrumentation Module}
AFL comes with a special compiler pass for \texttt{clang} that instruments every branch instruction to enable coverage feedback. By extending this compiler, we instrument a program under test  at three levels: specific locations, basic blocks, and the application.

\paragraph*{Specific locations.} \tool takes a list of program locations at which program behaviors associated with a property under test might occur. At each of the given program locations, the instrumentation module injects two components: {\em event generator} and {\em state recorder}. Event generator is a piece of code that generates an event when the provided condition is satisfied at run-time. The state recorder is a component that takes a snapshot of program states and generates a hash code for the state when the given program location is reached in the execution. 

\paragraph*{Basic blocks.} \tool guides fuzzing to a target using the feedback on how close to the target an input is as explained in Section~\ref{sec:guided-fuzzing}. At runtime, \tool requires the distance from each basic block to the target on the CFG (control flow graph). The instrumentor instruments a function call in each basic block at runtime. The function call will query a table that stores distances from each block to program locations associated with the given property (i.e., targets). The distance from a basic block to each program location is computed offline with the distance calculator component that is borrowed from AFLGo \cite{aflgo}.

\paragraph*{Applications.} For a program under test, the instrumentation module injects a {\em monitor} into the program. During fuzzing, the monitor collects events generated by instrumented event generators and produces execution traces. For property checking, the monitor leverages Spot libraries~\cite{spot} to generate a B{\"u}chi automata from the negation of an LTL property and validates these traces. The instrumentation module also instruments an {\em observer} in the program that monitors execution of inputs; it maps a given suitable execution trace prefix to the input event sequence producing it, so that the occurrence of the prefix can be detected by the observer, during fuzzing. The fuzzing process then seeks to further extend this prefix with "suitable" events as described in the following.

\subsection{Fuzzer}
\begin{figure}[t]
    \centering
    \includegraphics[width=0.38\textwidth]{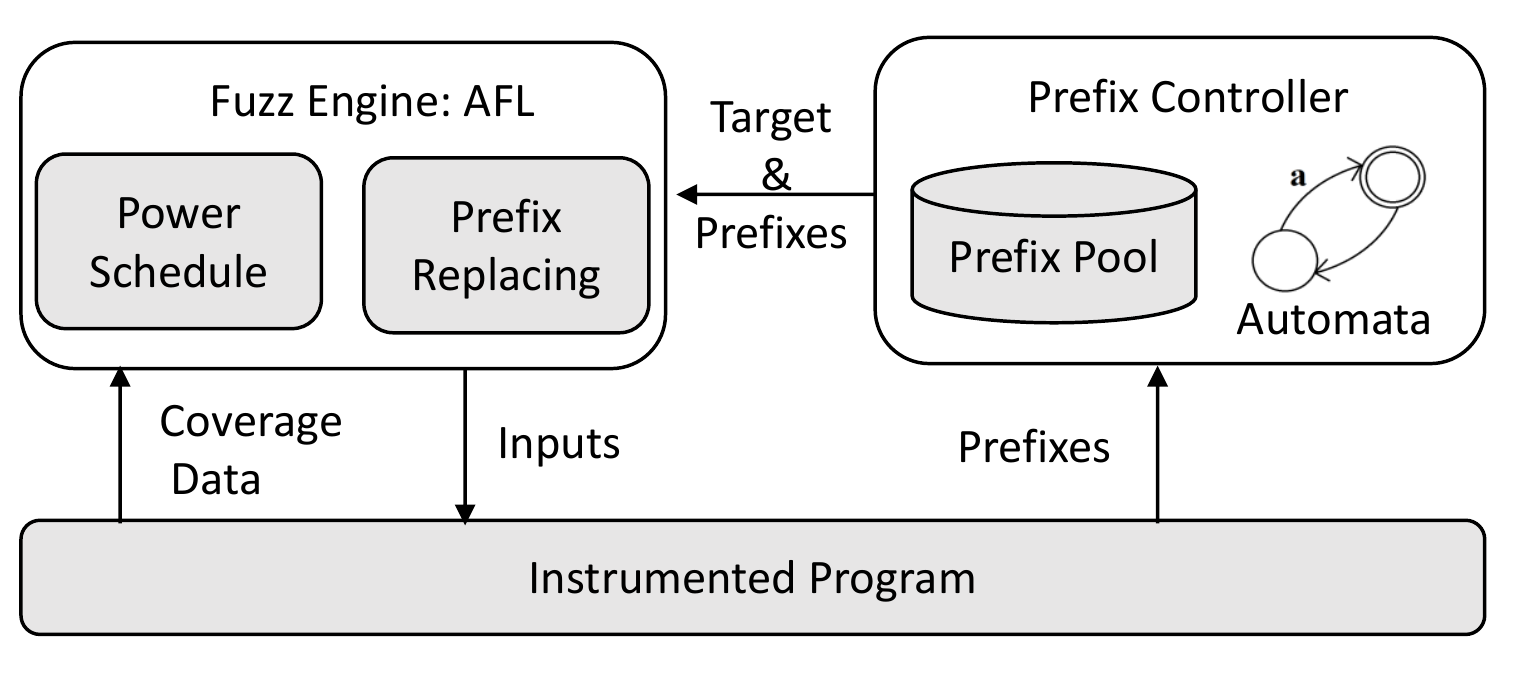}
    \vspace*{-0.1in}
    \caption{The architecture of \toolBold.}
    \label{fig:arch}
    \vspace*{-0.1in}
\end{figure}

Figure~\ref{fig:arch} shows the Fuzzer component's architecture. It mainly comprises two modules: {\em prefix controller} and {\em fuzz engine}. \tool saves input prefixes whose execution traces make transitions on the automata and reuses them for further exploration (Section~\ref{sec:guided-fuzzing}). At runtime, the prefix controller conducts three tasks: (1) collecting prefixes reported by the monitor instrumented in the program under test and storing them into a pool; (2) selecting a prefix from the pool for further exploration according to Algorithm~\ref{alg:fuzzing}; (3) identifying the target program location based on the selected prefix. The fuzz engine is obtained by modifying AFL \cite{afl}. It generates inputs starting with a given input prefix. To reach a target, our fuzzer integrates the power scheduling component developed in AFLGo \cite{aflgo} to direct fuzzing. In \tool, we direct execution to reach a target after the execution of an input prefix. Thus, the fuzz engine collects no feedback, such as coverage data during execution of the input prefix, and only collects feedback data after the execution of the input prefix is completed. 
\section{Evaluation\label{sec:evaluation}}

In our experiments, we seek to answer the following questions:
\begin{itemize}
    \item [\textbf{RQ1}] \textbf{Effectiveness:} How effective is \tool at finding LTL property violations?  
    \item [\textbf{RQ2}] \textbf{Comparison:} How does \tool compare to the state-of-the-art validation tools in terms of finding LTL property violations?
    \item [\textbf{RQ3}] \textbf{Usefulness:} How useful is \tool in revealing LTL property violations in real-world systems?   
\end{itemize}

\subsection{Subject Programs}

Table~\ref{tab:subjects} lists the subject programs used in our evaluation.  This includes 7 open source software projects that implement 6 widely-used network protocols. We selected these projects because they (1) are reactive software systems that \tool is designed for, (2) include appropriate specification documents from which LTL properties can be generated, and (3) are widely-used and have been studied. Finding bugs in such real-world systems is thus valuable.
\begin{table}[h]
\small
\caption{Detailed information about our subject programs.}
\setlength{\abovecaptionskip}{3pt}%
\setlength{\belowcaptionskip}{-10.9cm}%
\setlength{\tabcolsep}{2.5pt}
\label{tab:subjects}
\begin{tabular}{l|crcc}
\toprule
\textbf{Project} & \textbf{Protocol} & \textbf{\#SLOC} & \textbf{InPreviousWork} & \textbf{GithubStars} \\
 \hline
ProFTPD~\cite{proftp} & FTP & 210.8k & \cite{profuzzbench}  & 339 \\ 
Pure-FTPd~\cite{pureftp} & FTP & 52.9k & \cite{profuzzbench} & 435 \\
Live555~\cite{live555} & RTSP & 52.5k & \cite{aflnet} \cite{profuzzbench} & 526 \\
OpenSSL~\cite{openssl} & TLS & 286.7k & \cite{chiron} \cite{profuzzbench} \cite{tlschecker} & 16.3K \\
OpenSSH~\cite{openssh} & SSH & 98.3k & \cite{sshchecker} \cite{profuzzbench} & 1.5K \\
TinyDTLS~\cite{tinydtls} & DTLS & 63.2k & \cite{dtlschecker} \cite{profuzzbench} & 43   \\
Contiki-Telnet~\cite{telnet} & TELNET & 353.4k & \cite{chiron} & 3.4K \\
\bottomrule
\end{tabular}
\end{table}

\subsection{Experiment Setup}
To answer the research questions, we conducted three empirical studies on the subject programs. 

\begin{table*}[]
    \caption{Statistics of found violations and the performance of four tools in finding the violations.}
    \label{tab:reproduction_results}
    \small
    \setlength{\abovecaptionskip}{0pt}%
    \setlength{\belowcaptionskip}{-12pt}%
    \setlength\tabcolsep{4pt}
    \def\arraystretch{1.1}
    \begin{threeparttable}
    \begin{tabular}{l|ll|ll|r|rr|rr|rr}
        \toprule
        \multirow{2}{*}{\bfseries Prop} & \multirow{2}{*}{\bfseries CVE-ID} & \multirow{2}{*}{\bfseries Type of Vulnerability} & \multirow{2}{*}{\bfseries Program} & \multirow{2}{*}{\bfseries Version} & \multicolumn{1}{c|}{\toolBold} & \multicolumn{2}{c|}{\paflgoBold} & \multicolumn{2}{c|}{\aflgoBold} & \multicolumn{2}{c}{\checkerBold} \\ 
        \cline{6-12} 
         &  &  &  &  & {\bfseries Time(h)} & {\bfseries Time(h)} & $\hat{A}_{12}$ & {\bfseries Time(h)} & $\hat{A}_{12}$ & {\bfseries Time(h)} & $\hat{A}_{12}$ \\
        \hline
        \hline
        $PrF_1$ & CVE-2019-18217 & Infinite Loop & ProFTPD & 1.3.6 & 4.62 & T/O & $\textbf{1.00}$ & T/O & $\textbf{1.00}$ & T/O & $\textbf{1.00}$ \\
        $PrF_2$ & CVE-2019-12815 & {Illegal File Copy} & ProFTPD & 1.3.5 & 0.95 & 2.01 & $\textbf{0.84}$ & {T/O} & $\textbf{1.00}$ & T/O & $\textbf{1.00}$ \\
        $PrF_3$ & CVE-2015-3306 &  Improper Access Control & ProFTPD & 1.3.5 & 1.14 & 1.89 & \textbf{0.76} & T/O & $\textbf{1.00}$ & T/O & $\textbf{1.00}$ \\
        $PrF_4$ & CVE-2010-3867 &  {Illegal Path Traversal} & ProFTPD & 1.3.3 & 2.06 & 5.17 & $\textbf{0.85}$ & T/O & $\textbf{1.00}$ & T/O & $\textbf{1.00}$ \\
        \hline
        $LV_1$ & CVE-2019-6256 &  {Improper Condition Handle} & Live555 & 2018.10.17 & 5.29 & 11.13 & $\textbf{1.00}$ &  11.47 & $\textbf{1.00}$ & T/O & $\textbf{1.00}$ \\
        $LV_2$ & CVE-2019-15232 & {Use after Free} & Live555 & 2019.02.03 & 0.22 &  1.42 & $\textbf{0.91}$ &  1.46 & $\textbf{0.92}$ & T/O & $\textbf{1.00}$ \\
        $LV_3$ & CVE-2019-7314 & {Use after Free} & Live555 & 2018.08.26 & 1.27 &  4.18 & $\textbf{0.98}$ &  T/O & $\textbf{1.00}$ & T/O & $\textbf{1.00}$ \\
        $LV_4$ & CVE-2013-6934 & {Numeric Errors} & Live555 & 2013.11.26 & 2.73 &  2.58 & 0.40 & 2.21 & 0.39 & T/O & $\textbf{1.00}$ \\
        $LV_5$ & CVE-2013-6933 & {Improper Operation Limit} & Live555 & 2011.12.23 &  1.80 &  1.99 & 0.63 &  1.45 & 0.33 & T/O & $\textbf{1.00}$ \\
        \hline
        $SH_1$ & CVE-2018-15473 & {User Enumeration} & OpenSSH & 7.7p1 & 0.18 & 0.17 & 0.44 & T/O & $\textbf{1.00}$ & 24.00 & $\textbf{1.00}$ \\
        $SH_2$ & CVE-2016-6210 & {User Information Exposure} & OpenSSH & 7.2p2 & 0.19 & 0.19 & 0.50 & T/O & $\textbf{1.00}$ & 24.00 &  $\textbf{1.00}$ \\
        \hline
        $SL_1$ & CVE-2016-6309 & {Use after Free} & OpenSSL & 1.1.0a & 3.77 & 6.00 & $\textbf{0.74}$ &  6.58 & $\textbf{0.82}$ & T/O & $\textbf{1.00}$ \\
        $SL_2$ & CVE-2016-6305 & Infinite Loop & OpenSSL & 1.1.0 & 1.45 & T/O & $\textbf{1.00}$ & T/O & $\textbf{1.00}$ & T/O & $\textbf{1.00}$ \\
        $SL_3$ & CVE-2014-0160 & Illegal Memory Access & OpenSSL & 1.0.1f & 1.11 & 7.31 & $\textbf{1.00}$ & T/O & $\textbf{1.00}$ & T/O & $\textbf{1.00}$ \\
        \hline
        \hline
        \multicolumn{3}{l|}{Found violations in total} &
        \multicolumn{2}{c|}{-} & 14 & \multicolumn{2}{r|}{12} & \multicolumn{2}{r|}{5} & \multicolumn{2}{r}{2} \\
        \hline
        \multicolumn{3}{l|}{Average time usage (hours)} &
        \multicolumn{2}{c|}{-} & 1.91 & \multicolumn{2}{r|}{6.57} & \multicolumn{2}{r|}{17.08} & \multicolumn{2}{r}{24.00} \\
        \hline
        \multicolumn{3}{l|}{Comparison with \tool on time usage} & 
        \multicolumn{2}{c|}{-} & - & \multicolumn{2}{r|}{3.44x} & \multicolumn{2}{r|}{8.93x} & \multicolumn{2}{r}{12.55x} \\
        \bottomrule
        \end{tabular}
        \begin{tablenotes}
            \footnotesize
            \item[1] T/O represents tools cannot expose vulnerabilities within 24 hours for 10 experimental runs. We replace T/O with 24 hours when calculating average usage time.
            \item[2] Statistically significant values of $\hat{A}_{12}$ are shown in bold.
        \end{tablenotes}
    \end{threeparttable}
\end{table*}                     

\subsubsection{Effectiveness of \tool.} We evaluate \tool's effectiveness by running it on a set of LTL properties in subject programs where violations are already known; we check the number of LTL properties for which \tool can find violations. To create such a dataset, we collect event ordering related CVEs (so that they can be captured as a temporal property)
that are disclosed in subject programs, e.g., an FTP client copies files from the server without logging in successfully.  Specifically, for each subject, we select 10 such CVEs with criteria: (1) reported recently (during 2010-2020);  (2) include instructions to reproduce the bug, (3) relevant to event orderings.  Then we manually reproduce them with the corresponding version of code. If a CVE is reproducible,  then we write the property in LTL and put it in our dataset of LTL properties. Based on the aforementioned criteria, we collected 14 CVEs in 7 subjects as shown in Table~\ref{tab:reproduction_results}; these LTL properties can be found in our dataset \footnote{\color{blue}{https://github.com/ltlfuzzer/LTL-Fuzzer/tree/main/ltl-property}} and the appendix of this paper. Our goal is to check experimentally if \tool can find violations of these  LTL properties.

\subsubsection{Comparison with other tools.} We evaluate \tool and state-of-the-art techniques on the LTL property  above and compare them in terms of the number of LTL properties for which each technique finds the violations and the time that is used to find a violation. For state-of-the-art techniques, we reviewed recent and well-known techniques in model checking, runtime verification and directed fuzzing domains. We chose the following techniques for comparison with \tool.
\begin{itemize}[leftmargin=*]
    \item \aflgo~\cite{aflgo}. It is a well known directed greybox fuzzer which drives execution to a target with  a simulated annealing-based power schedule that assigns more energy to inputs that hold the trace closer to the target. We take it as a baseline tool.
    \item \paflgo. It is an implementation which enables \aflgo to detect an LTL property violation. Specifically, \paflgo powers \aflgo with only the LTL test oracle such that it can report an error when the given LTL property is violated in the execution. By comparing with \paflgo, we evaluate how effective is our automata-guided fuzzing strategy in finding LTL property violations. Note that \paflgo is also a tool built by us, but it lacks the automata guided fuzzing of \tool.  
    \item \checker. It  combines model learning and model checking to verify properties in a software system. Specifically, it leverages a learning library called LearnLib~\cite{learnlib} to build a model for the software system and then verifies given properties on the learned model with the well-known model checker NuSMV~\cite{nusmv}. In the paper, we indicate it with \checker. This technique was published at CAV 2016 \cite{tcpchecker} and has been subsequently adopted in recent works such as \cite{MPINSPECTOR} and \cite{dtlschecker}. 
\end{itemize}
We briefly summarize why we did not include certain other model-checkers and fuzzers, and all runtime verification tools for comparison. Model checking tools CBMC~\cite{cbmc}, CPAChecker~\cite{cpachecker}\footnote{For some tools, the LTL checker module is not available for usage / experimentation, as our email enquiry with CPAchecker team revealed.}, Seahorn~\cite{seahorn}, SMACK~\cite{smack}, UAutomizer~\cite{automizer}, DIVINE~\cite{divine} cannot support LTL property verification. Schemmel's work~\cite{loop:cav} published at CAV 2018 partially supports LTL property verification. SPIN~\cite{spin} supports LTL property verification but only works with a modeling language Promela~\cite{promela} and the tool provided in SPIN for extracting models from C programs failed to work on our subject programs. Some model checking tools~\cite{chiron, codethorn}, and directed fuzzing tools  (like UAFL~\cite{uafl},  Hawkeye~\cite{hawkeye} and TOFU~\cite{tofu}) we reviewed, are not publicly available. 

Finally, all of available runtime verification tools \cite{taxonomyRV} (like JavaMOP \cite{javamop}, MarQ \cite{marq} and Mufin \cite{muffin}) cannot check LTL properties in C/C++ software systems. Furthermore, our method is conceptually different and complementary to runtime verification --- our method generates test executions, while runtime verification checks a test execution. While the combination of our method with runtime verification is possible, a comparison is less meaningful.  

\subsubsection{Real-world utilty.} In this study, we read RFC specifications that these subject programs follow to extract temporal properties and describe them in LTL. Then we use \tool to check these properties on the subject programs. 

\paragraph{Configuration Parameters.}Following fuzzing evaluation suggestions from the community~\cite{kleesevaluating}, we run each technique for 24 hours and repeat each experiment 10 times to achieve statistically significant results. For the initial seeds, we use seed inputs provided in ProFuzzBench~\cite{profuzzbench} for all subjects. ProFuzzBench is a benchmark for stateful fuzzing of network protocols, which contains a suite of representative open-source network protocol implementations. For  Contiki-Telnet, which is not contained in ProFuzzBench, we generate random inputs as its initial seeds. For \tool, we need to specify the time budget for reaching a single program location and we configure it with 45 minutes for each target. For \aflgo and \paflgo, we need to provide a target for an LTL property being checked. We specify the target by randomly selecting from program locations that are associated with atomic propositions that trigger the transition to an accepting state on the automata of the negation of the property.  In the example in Section~\ref{sec:approach}, we chose one of loop entries as the target since proposition $o$ triggers the transition to the accepting state shown in Figure~\ref{fig:automata} and it corresponds with loop entries. For execution environments, we conducted experiments on a physical machine with 64 GB RAM and a 56 cores Intel(R) Xeon(R) E5-2660 v4 CPU, running a 64-bit Ubuntu TLS 18.04 as the operating system.

\subsection{Experimental Results}


\begin{table*}[t]
\small
\caption{Zero-day Bugs found by \toolBold; for several of them CVEs have been assigned but CVE ids are not shown.}
\setlength{\abovecaptionskip}{3pt}%
\setlength{\belowcaptionskip}{6pt}%
\label{tab:violation}
\begin{tabular}{m{0.5cm}|m{2.2cm}|m{11.6cm}|m{2cm}}  
\toprule
 \textbf{Prop} & \textbf{Project} & \textbf{Description of violated properties} & \textbf{Bug Status} \\
 \hline
 \hline
 $TD_1$ & TinyDTLS (0.9-rc1) &{If the server is in the \texttt{WAIT\_CLIENTHELLO} state and receives a \texttt{ClientHello} request with valid cookie and the epoch value 0, must finally give \texttt{ServerHello} responses.} & CVE-2021-42143, Fixed \\
 \hline
 $TD_2$ & TinyDTLS (0.9-rc1) &If the server is in \texttt{WAIT\_CLIENTHELLO} state and receives a \texttt{ClientHello} request with valid cookie but not 0 epoch value, must not give
\texttt{ServerHello} responses before receiving \texttt{ClientHello} with 0 epoch value. & CVE-2021-42142, Fixed \\

\hline
$TD_{3}$ & TinyDTLS (0.9-rc1) & {If the server is in the \texttt{WAIT\_CLIENTHELLO} state and receives a \texttt{ClientHello} request with an invalid cookie, must reply \texttt{HelloVerifyRequest}.} & CVE-2021-42147, Fixed  \\
\hline
$TD_{5}$ & TinyDTLS (0.9-rc1) & {If the server is in the \texttt{DTLS\_HT\_CERTIFICATE\_REQUEST} state and receives a \texttt{Certificate} request, must give a  \texttt{DTLS\_ALERT\_HANDSHAKE\_FAILURE} or \texttt{DTLS\_ALERT\_DECODE\_ERROR} response, or set \texttt{Client\_Auth} to be verified.}  & CVE-2021-42145, Fixed \\
\hline
$TD_{11}$ & TinyDTLS (0.9-rc1) & {After the server receives a \texttt{ClientHello} request without renegotiation extension and gives a \texttt{ServerHello} response, then receives a \texttt{ClientHello} again, must refuse the renegotiation with an Alert.} & Confirmed\\
\hline
$TD_{12}$ & TinyDTLS (0.9-rc1) & {After the server receives a \texttt{ClientHello} request and gives a \texttt{ServerHello} response, then receives a \texttt{ClientKeyExchange} request with a different epoch value than that of \texttt{ClientHello}, server must not give \texttt{ChangeCipherSpec} responses.} & CVE-2021-42141, Fixed \\
\hline
$TD_{13}$ & TinyDTLS (0.9-rc1) & {After the server receives a \texttt{ClientHello} request and gives a \texttt{ServerHello} response, then receives a \texttt{ClientHello} request with the same epoch value as that of the first one, server must not give \texttt{ServerHello}.} &  CVE-2021-42146 \\
\hline
$TD_{14}$ & TinyDTLS (0.9-rc1) & If the server receives a \texttt{ClientHello} request and gives a \texttt{HelloVerifyRequest} response, and then receives a over-large packet even with valid cookies, the server must refuse it with an \texttt{Alert}.  &  CVE-2021-42144, Fixed  \\
\hline
$CT_{1}$ & Contiki-Telnet (3.0)  & After \texttt{WILL} request is received and the corresponding option is disabled, must send \texttt{DO} or \texttt{DONT} responses. & CVE-2021-40523\\
\hline
$CT_{2}$ & Contiki-Telnet (3.0)  & After \texttt{DO} request is received and the corresponding option is disabled, must send \texttt{WILL} or \texttt{WONT} responses. & Confirmed\\
\hline
$CT_{7}$ & Contiki-Telnet (3.0)  & After \texttt{WONT} request is received and the corresponding option is disabled, must not give responses. & CVE-2021-38311 \\
\hline
$CT_{8}$ &Contiki-Telnet (3.0)  & After \texttt{DONT} request is received and the corresponding option is disabled, must not give responses. & Confirmed \\
\hline
$CT_{10}$ &Contiki-Telnet (3.0)  & Before \texttt{Disconnection}, must send an Alert to disconnect with clients. & CVE-2021-38387 \\
\hline
$CT_{11}$ &Contiki-Telnet (3.0)  & If conducting \texttt{COMMAND} without \texttt{AbortOutput}, the response must be same as the real execution results. & CVE-2021-38386 \\
\hline
$PuF_5$ &Pure-FTPd (1.0.49)  & When quota mechanism is activated and user quota is exceeded,  must finally reply a quota exceed message. & CVE-2021-40524, Fixed \\
\bottomrule
\end{tabular}
\end{table*}

\subsubsection{[RQ1] Effectiveness.} \label{sec:question1}
Table~\ref{tab:reproduction_results} shows property violations found by \tool for  the 14 LTL properties derived from known CVEs. The first column shows identifiers of the properties being checked. The corresponding LTL properties and their descriptions can be found in our dataset. Columns 2 - 5 represent CVE-IDs, types of vulnerabilities that CVEs represent, subject names, and subject versions, respectively. Column ``\tool'' shows the time that is used to find a violation by \tool. As shown in Table~\ref{tab:reproduction_results}, \tool can effectively detect violations of LTL properties in the subjects. It successfully detected the violation for all the 14 LTL properties in the dataset. On average, it took \tool 1.91 hours to find a violation.  

\result{\tool is found to be effective in finding LTL property violations, detecting violations for all 14 properties derived from known CVEs.}

\subsubsection{[RQ2] Comparison.} \label{sec:question2}
As shown in Table~\ref{tab:reproduction_results}, the last three main columns show the time that is used for comparison techniques to find a violation on the 14 LTL properties in the experiment. Note that ``T/O'' indicates a technique failed to find the violation for an LTL property in the given time budget (i.e., 24 hours). To mitigate randomness in fuzzing, we adopted the Vargha-Delaney statistic  $\hat{A}_{12}$~\cite{a12} to evaluate whether one tool significantly outperforms another in terms of the time that is used to find a violation.  The $\hat{A}_{12}$ is a non-parametric measure of effect size and gives the  probability that a randomly chosen value from data group 1 is  higher or lower than one from data group 2. It is commonly used to evaluate whether the difference between two groups of data is significant. Moreover, we also
use Mann-Whitney U test to measure the statistical significance of performance gain. When it is significant (taking 0.05 as a significance level), we mark the $\hat{A}_{12}$ values in bold. 

\tool found violations of all of the 14 LTL properties, followed by \paflgo (12), \aflgo (5), and \checker (2). We note that \paflgo is also a tool built by us, it partially embodies the ideas in \tool and is meant to help us understand the benefits of automata-guided fuzzing. In terms of the time that is used to find a violation, \tool is the fastest (1.91 hours), followed by \paflgo (6.57 hours), \aflgo (17.08 hours), and \checker (24.00 hours). In other words, \tool is 3.44x, 8.93x, 12.55x faster than \paflgo, \aflgo, and \checker, respectively.  For CVE-2013-6934 and CVE-2013-6933, \aflgo performed slightly better than other techniques, while \paflgo exhibited the same performance as \tool for CVE-2018-15473 and CVE-2016-6210. We investigated these 4 CVEs and found that triggering those vulnerabilities is relatively straightforward. They can be triggered without sophisticated directing strategies. As a result, other techniques achieve a slightly better performance than \tool for these four CVEs.  In terms of the $\hat{A}_{12}$ statistic, \tool performs significantly better than other techniques in most cases.   
\result{ \tool found violations of all the 14 LTL properties in the experiment. \paflgo, \aflgo and \checker found 12, 5, 2 property violations, respectively. \tool is 3.44x, 8.93x, 12.55x faster than \paflgo, \aflgo, and \checker.}

\subsubsection{[RQ3] Real-world utility.} \label{sec:question3} 

 In this study, we evaluate utility of \tool by checking whether it can find zero-day bugs in  real-world protocol implementations.
We extract 50 properties from RFCs that our subject programs follow (aided by comments in the source code of the programs) and write them in linear-time temporal logic. The details of the 50 LTL properties can be found in our dataset.
In the experiment, \tool achieved a promising result. Out of these 50 LTL properties, \tool discovered {\em new} violations for 15 properties, which are shown in Table~\ref{tab:violation}. We reported these 15 zero-day bugs to developers and all of them got confirmed by developers. We reported them on the common vulnerabilities and exposures (CVE) system (see https://cve.mitre.org/) and 12 of them were assigned CVE IDs. Out of 15 reported violations, 7 have been fixed at the time of the submission of our paper. Notably, \tool shows effectiveness in finding violations for liveness properties. In the experiment, \tool successfully found violations for 4 liveness properties which are $PrF_1$, $SL_2$, $TD_1$ and $PuF_5$. All the 4 violations were confirmed by developers, i.e., they are not spurious results. Moreover, to discover violations for these 4 liveness properties, \tool only recorded 6, 11, 4 and 9 states, respectively. Since every state is 
recorded as a 32-bit integer, the memory consumption for recording states is thus found to be negligible in our experiments.

\result{Among 50 LTL properties extracted from protocol RFCs, \tool found 15 previously unknown violations in  protocol implementations and 12 of these have been assigned CVEs.}

\subsection{Threats to validity}
There are potential threats to validity of our experimental results. One concern is \textit{external validity}, i.e., the degree to which our results can be generalized to and across other subjects. To mitigate this concern, we selected protocol implementations that are widely used and have been frequently evaluated in previous research (as shown in Table~\ref{tab:subjects}). We may have made mistakes in converting informal requirements into LTL properties. To reduce this kind of bias, we let two authors check generated properties and remove those on which they do not agree, or do not think are important properties. 

In principle, \tool can report false positives due to incorrect instrumentation, e.g., if we fail to instrument some target locations for an atomic proposition. We mitigate the risk of false positives by checking the reported counterexamples and validating that they are true violations of the temporal property being checked. We add here that we did not encounter such false positives in any of our experiments.

Another concern is \textit{internal validity}, i.e., the degree to which our results minimize systematic error. First, to mitigate spurious observations due to the randomness in the fuzzers and to gain statistical significance, we repeated each experiment 10 times and reported the Vargha-Delaney statistic $\hat{A}_{12}$. Secondly, our \tool implementation may contain errors. To facilitate scrutiny, we make \tool code available. 

\section{Related Work} \label{sec:relatedWork}







\paragraph*{Model Checkers}
Model checking is a well-known property verification technique dating back to 1980s \cite{Clarke1981,Sifakis}; it is used to prove a temporal property in a finite state system, or to find property violation bugs. The early works check a temporal logic property against a finite state transition system. 
There exist well-known model checkers such as ~\cite{spin,nusmv,ltsmin} which can be used to check temporal properties on a  constructed model (via state space exploration). To construct models, one method is manual construction via a modeling language. This requires substantial effort and can be error-prone~\cite{cmc, practical}.
\tool directly checks software implementations; it does not separately extract models from software.

Early works on model checking  have been extended to automatically find bugs in software systems, which are typically infinite-state systems.
Model checking of software systems usually involves either some extraction of finite state models, or directly analyzing the infinite state software system via techniques such as symbolic analysis.  Automatic model extraction approaches \cite{bandera, predicate, slicing, inference} include the works on predicate abstraction and abstraction refinement \cite{predicate,slam} which build up a hierarchy of finite-state abstract models for a software system for proving a property. These approaches extract models which are conservative approximations and capture a superset of the program behavior. 
There are a number of stateful software model checkers, such as CMC~\cite{cmc}, Java Pathfinder~\cite{javapathfinder}, MaceMC~\cite{macemc}, CBMC~\cite{cbmc}, CPAchecker~\cite{cpachecker}, which find assertion violations in software implementations. Many of these checkers do not check arbitrary LTL properties for software implementations.
These model checkers either suffer from state space explosion, or suffer from other kinds of explosion such as the explosion in the size/solving-time for the logical formula in bounded model checking. In contrast, \tool does not save any states for safety property checking and saves only certain property-relevant program states in liveness property checking. At the same time, \tool does not give verification guarantees and does not perform complete exploration of the state space. We now proceed to discuss incomplete validation approaches.

\paragraph*{Incomplete Checkers}
Instead of exploring the complete set of behaviors, or a super-set of behaviors, one can also explore a subset of behaviors. Incomplete model learning approaches \cite{modellearning} can be mentioned in this regard.  The active model learning technique, such as LearnLib~\cite{learnlib}, is widely used to learn models of real-world protocol implementations~\cite{dtlschecker, tlschecker, sshchecker, tcpchecker}. It does not need user involvement. But it is time-consuming and hard to determine whether the learned model represents the complete behavior of the software system~\cite{modellearning, combinedlearning}. Compared with the active learning, \tool can more rapidly check properties, as shown in our experimental comparison with LearnLib+NuSMV. 
To alleviate the state-explosion problem, stateless checkers such as VeriSoft~\cite{verisoft} and Chess~\cite{chess} have been proposed; these checkers do not store program states. These works typically involve specific search strategies to check {\em specific} classes of properties such as deadlocks, assertions and so on. In contrast, \tool represents a general approach to find violations of {\em any} LTL property.

\paragraph*{Runtime Verification} 
Runtime verification is a lightweight and yet rigorous verification technique \cite{briefRV, introductionRV}. It analyzes a single execution trace of a system against formally specified properties (e.g., LTL properties). It origins from model checking and applies model checking directly to the real implementations. Model checking checks a model of a target system to verify correctness of the system, while runtime verification directly checks the implementation, which could avoid different behaviours between models and implementations. \tool shares the same benefit as runtime verification. Besides, runtime verification deals with finite executions, as one single execution has necessarily to be finite. This avoids the state explosion problem that model checking suffers from. Meanwhile, it leads to that runtime verification approaches \cite{mop, javamop, marq, muffin} often only check safety properties. \tool, however, is able to check liveness properties by leveraging the strategy of saving program states. 

Conceptually, our method is very different from runtime verification. Runtime verification focuses on the checking (a temporal logic property) on a single execution. Our method is focused on using temporal logic property to guide the construction of an execution which violates the property. Thus our work is more of a test generation method. Since runtime verification methods need tests whose execution will be checked, our method can be complementary to runtime verification. In other words, our method can generate tests likely to violate a given temporal property, and these tests can further validated by run-time verification. 


\paragraph*{Greybox Fuzzing}
There are three broad variants of fuzzing: blackbox fuzzing~\cite{blackboxfuzzing}, whitebox fuzzing or symbolic execution~\cite{klee, loop:cav, chiron}, and greybox fuzzing~\cite{afl, bohme2017coverage, rawat2017vuzzer, aflgo, hawkeye, uafl}. We first discuss greybox fuzzing since they are the most widely used in industry today.
In contrast to software model checking, blackbox/greybox fuzzing techniques represent a random/biased-random search over the domain of inputs for finding bugs or vulnerabilities in programs.
Most greybox fuzzing techniques are used to detect memory issues (e.g., buffer overflow and use after free) that can produce observable behaviors (e.g., crashes). However, \tool can not only witness simple properties like memory corruption, but also detect LTL property violations, for {\em any} given LTL property, however complex. Recent advances in greybox fuzzing use innovative objective functions for achieving different goals, such as \cite{aflgo} directs the search to specific program locations. The capabilities of \tool go beyond visiting specific locations, and \tool is used to witness specific event ordering constraints embodied by the negation of an arbitrary LTL property. PGFUZZ~\cite{pgfuzz} is a greybox-fuzzing framework to find safety violations for robotic vehicles, but it is customized to be used on implementations of robotic vehicles. \tool can be used to find violation of {\em any} LTL property for software from any application domain.

\paragraph*{Symbolic Execution based Validation}
Symbolic execution or whitebox fuzzing approaches are typically used to find violations of simple properties such as assertions \cite{dart,klee}. Recent whitebox fuzzing techniques do find violations of certain classes of properties. Schemmel's work~\cite{loop:cav} checks liveness properties while CHIRON~\cite{chiron} checks safety properties. \cite{regularproperty1, regularproperty2} proposed regular-property
guided dynamic symbolic execution to find the program paths satisfying a property.  However, all of these approaches  require a long time budget for heavy-weight program analysis and back-end constraint solving. As a result, these techniques face challenges in scalability. In contrast, \tool is built on top of greybox fuzzing; it can validate  arbitrarily large and complex software implementations. 
\section{Perspective\label{sec:conclusion}}

We present \tool, a linear-time temporal logic guided grey-box fuzzing technique, which takes Linear-time Temporal Logic (LTL) properties extracted from informal requirements such as RFCs and finds violations of these properties in C/C++ software implementations. Our evaluation shows that \tool is effective in finding property violations. It detected 15 LTL property violations in real world protocol implementations that were previously unknown; 12 of these zero day bugs have been assigned CVEs. We make the data-set of LTL properties, and our tool available for scrutiny.

Our work shows the promise of synergising concepts from temporal property checking  with recent advances in greybox fuzzing (these advances have made greybox fuzzing more systematic and effective). Specifically, in this paper we have taken concepts from automata-theoretic model checking of LTL properties \cite{Vardi86}, while at the same time adapting/ augmenting directed greybox fuzzing \cite{aflgo}. The main advancement of greybox fuzzing in our work, is the ability to find violations of {\em arbitrary} LTL properties, which is achieved by borrowing the B{\"u}chi automata construction from \cite{Vardi86}.
We note that the real-life practical value addition of software model checking is often from automated bug-finding in software implementations rather than from formal verification. Runtime verification complements software model checking by analyzing a single execution trace of software implementations. Our work essentially shows the promise of enjoying the main practical benefits of software model checking more efficiently and effectively via augmentation of (directed) greybox fuzzing. This is partially shown by the experiments in this paper where we have compared our work with both model checkers and fuzzers. Our work is also complementary to runtime verification since we generate test executions guided by a LTL property, while runtime verification would check a LTL property against a single test execution.

Arguably we could compare \tool with more model checkers and fuzzers, experimentally. At the same time, we have noted that many model checkers were found to be not applicable for checking arbitrary LTL properties of arbitrary C/C++ software implementations. Moreover, the problem addressed by \tool is certainly beyond the reach of fuzzers since fuzzers cannot detect temporal property violations. Overall, we believe our work represents a practical advance over model checkers and runtime verification, and a conceptual advance over greybox fuzzers. We expect that the research community will take the work in our paper forward, to further understand the synergies among software model checking, runtime verification and greybox fuzzing.

\section*{ACKNOWLEDGEMENTS}
{This work was partially supported by the National Satellite of Excellence in Trustworthy Software Systems and funded by National Research Foundation (NRF) Singapore under National Cybersecurity R\&D (NCR) programme.}

\balance
\bibliographystyle{ACM-Reference-Format}
\bibliography{bibliography}

\newpage
\appendix
\onecolumn

\section*{Appendix}

\section{Properties extracted from Existing CVEs} \label{sec:cve_property}

 
\begin{center}
\small 
\setlength{\abovecaptionskip}{3pt}%
\setlength\tabcolsep{2pt}
\setstretch{1.05}
\label{tab:cve_description}
\begin{longtable}{m{0.8cm}|m{2.3cm}|m{7.9cm}|m{6.8cm}}
\caption{Properties extracted from existing CVEs in the implementations of Pro-FTPD ($PrF$) for the FTP protocol, Live555 ($LV$) for the RTSP protocol, OpenSSH ($SH$) for the SSH protocol, OpenSSL ($SL$) for the TLS protocol, TinyDTLS ($TD$) for the DTLS protocol, Contiki-Telnet ($CT$) for the TELNET protocol, and Pure-FTPd ($PuF$) for the FTP protocol.}   \\
\toprule
\textbf{ID} & \textbf{Vulnerability} & \textbf{Property Description} & \textbf{LTL Notation} \\
\hline
\hline
$PrF_1$ & CVE-2019-18217 & After one client succeeds to connect with a sever, the server should finally give responses for requests from the connected client. &	G((LogIN) $\rightarrow$ (X(G((Requests) $\rightarrow$ (X(F(Responses))))))) \\
\hline
$PrF_2$ & CVE-2019-12815	& If a client does not log in successfully, the server must not allow this client to copy files. &	G(($\neg$(LogIN)) $\rightarrow$ (X(G((request = CopyFiles) $\rightarrow$ X($\neg$(response = CopySuccessful))))))\\
\hline
$PrF_3$ & CVE-2015-3306  & If the server receives CPTO requests when the client doesn't succeed to log in, must not allow CPTO successfully. &	G(($\neg$(LogIN)) $\rightarrow$ (X(G((request = CPTO) $\rightarrow$ X($\neg$(response = CPTOSuccessful))))))\\
\hline
$PrF_4$ & CVE-2010-3867	& If the client logs in and is only assigned one writable directory, the server must not allow it to write out of scope of the assigned directory. &	G(((state = LogIN) $\wedge$ (WritableDirectory = true) $\wedge$ (request = OverWrite)) $\rightarrow$ (X(response = PermissionDenied))) \\
\hline
$LV_1$	& CVE-2019-6256 & After the connection channel is DESTROYED between the server and the client, the channel must not be USED unless one new connection is ESTABLISHED. &	 G((channel = DESTROYED) $\rightarrow$ (X(((channel = ESTABLISHED) R ($\neg$(channel = USED)))))) \\
\hline
$LV_2$	& CVE-2019-15232 & The server must not create two client sessions with the same ID. &	 G((SessionID = RID) $\rightarrow$ (X(G($\neg$(SessionID = RID))))) \\
\hline
$LV_3$	& CVE-2019-7314 & If the server receives the PLAY request in the INIT state, must not begin StartPlay & G(((state = INIT) $\wedge$ (request = PLAY)) $\rightarrow$ (X($\neg$(response = StartPlay))))\\
\hline
$LV_4$ & CVE-2013-6934 & If receiving a invalid request, must always refuse it with Method\_not\_Allowed. &  G((request = InvalidRequest) $\rightarrow$ X(response = Method\_not\_Allowed)) \\ 
\hline
$LV_5$ & CVE-2013-6933 & If receiving an invalid request, must always refuse it with Method\_not\_Allowed. & G((request = InvalidRequest) $\rightarrow$ X(response = Method\_not\_Allowed))  \\
\hline
$SH_1$ & CVE-2018-15473 & Whenever the server receives invalid username or valid username with wrong password, must give the same response. &  G(((request = InvalidUsername) $\vee$ (request = ValidUsername\&WrongPasswd)) $\rightarrow$ (X(G(SameResponse)))) \\
\hline
$SH_2$	& CVE-2016-6210 &  Whenever the server receives invalid username or valid username with wrong password, must give responses within the same time period. &	 G(((request = InvalidUsername) $\vee$ (request = ValidUsername\&WrongPasswd)) $\rightarrow$ (X(G(SameTimeToResponse)))) \\
\hline
$SL_1$	& CVE-2016-6309 & If the server receives the ChangeCipherSpec request after sending the ServerHello response, should give a ChangeCipherSpec response or an Alert. &	 G((response = ServerHello) $\rightarrow$ X((request = ChangeCipherSpec) $\rightarrow$ X((response = ChangeCipherSpec) $\vee$ (response = Alert)))) \\
\hline
$SL_2$	& CVE-2016-6305 & If the server receives an ApplicationData after the Handshake is successful, must finally give an Alert or response the ApplicationData. &  { G((state = HandshakeDone) $\wedge$ (request = ApplicationData)  $\rightarrow$  (X(F(response = ApplicationData)  $\vee$  (response = Alert)))))} \\
\hline
$SL_3$	& CVE-2014-0160 & After the server receives a ClientHello request with the Heartbeat\_extension in the peer\_allowed\_to\_send mode, and gives a ServerHello response with the same options, the sever receives a malformed Heartbeat request with the payload length field number larger than the real payload length, must always not send Heartbeat responses.  &  {G((((request = ClientHello) $\wedge$ (Heartbeat\_extension = true) $\wedge$ ( peer\_allowed\_to\_send = 1)) $\wedge$ X((response = ServerHello) $\wedge$ (Heartbeat\_extension = true) $\wedge$ (peer\_allowed\_to\_send = 1))) $\rightarrow$ F(G((((request = Heartbeat\_Request) $\wedge$ (Payload\_Length $\textgreater$ realPayloadLength))) $\rightarrow$ F(G($\neg$(response = Heartbeat\_Response))))))} \\
\bottomrule
\end{longtable}
\end{center} 

\section{LTL Properties extracted from RFC and Comments} \label{sec:rfc_property}

Please see the following pages for the Linear-time Temporal Logic properties extracted from sources such as RFCs.

\onecolumn

\begin{center}
\small 
\setlength{\abovecaptionskip}{3pt}%
\setstretch{1.05}
\label{tab:RFC_description}
\begin{longtable}{m{0.3cm}|m{0.6cm}|m{7.7cm}|m{8cm}} 
\caption{Properties extracted from relevant RFCs of network protocols and comments in the implementations of Pro-FTPD ($PrF$) for the FTP protocol, Live555 ($LV$) for the RTSP protocol, OpenSSH ($SH$) for the SSH protocol, OpenSSL ($SL$) for the TLS protocol, TinyDTLS ($TD$) for the DTLS protocol, Contiki-Telnet ($CT$) for the TELNET protocol, and Pure-FTPd ($PuF$) for the FTP protocol.} \\
\toprule
\textbf{NO} & \textbf{PID} & \textbf{Property Description} & \textbf{LTL Notation} \\
\hline
\hline
1 & $PrF_5$ & If receiving invalid username or invalid password, the server must always show the same message to the user. & G(((request = InvalidUsername) $\vee$ (request = InvalidPassword)) $\rightarrow$ X(G(sameResponse))) \\
\hline
2 & $PrF_6$ & If receiving the CWD request without login, the server must not give the CommandOkay response. & G(($\neg$(state = LogIN) $\wedge$ (request = CWD)) $\rightarrow$ X(G($\neg$(response = CommandOkay)))) \\
\hline
3 & $PrF_7$ & After a connection is constructed successfully, there should be a successful login and after that without failed login. & G(((request = ValidUserName\&ValidPasswd) $\rightarrow$ X(response = LoginSuccess)) $\rightarrow$ X(G($\neg$(response = LoginFailed))))  \\
\hline
4 & $PrF_8$ & After the connection is lost after a long time, responses should be always timeout. & G(LostConnection $\rightarrow$ X(G(response = Timeout)))\\
\hline
5 & $LV_6$ & If the server is in the Play state and receives a Pause request, should go into the Ready state. & G(((state = Play) $\wedge$ (request = Pause)) $\rightarrow$ X(state = Ready))  \\
\hline
6 & $LV_7$ & If the server is in the Play state and receives a TEARDOWN request, should go into the Init state. & G(((state = Play) $\wedge$ (request = TEARDOWN)) $\rightarrow$ X(state = Init)) \\
\hline
7 & $LV_8$ & If the server is in the Ready state and receives a Play request with one old URI, should response ChangeTransportParam. & G(((state = Ready) $\wedge$ (request = Play) $\wedge$ (OldURI = true)) $\rightarrow$ X(response = ChangeTransportParam)) \\
\hline
8 & $LV_9$ & If the server is connected with a client and then receives a TEARDOWN request, should finally give a TeardownSuccess or Timeout response. & G((((request = Setup) $\wedge$ X(response = SetupSuccess)) $\wedge$ X(request = TEARDOWN)) $\rightarrow$ X(F((response = TeardownSuccess) $\vee$ (response = Timeout))))\\
\hline
9 & $LV_{10}$ & The TEARDOWN request will not be acknowledged until the SETUP request is be acknowledged. &  G($\neg$((request = TEARDOWN) $\wedge$ X(response = TeardownSuccess)) U ((request = SETUP) $\wedge$ X(response = SetupSuccess)))  \\
\hline
10 & $SH_3$ & If the server receives the SSH\_MSG\_CHANNEL\_OPEN request and gives a SSH\_MSG\_CHANNEL\_OPEN\_ CONFIRMATION response, and then receives a Login request and gives a SSH\_MSG\_USERAUTH\_SUCCESS, there will not have a failure in user authentication.  & G(((request = SSH\_MSG\_CHANNEL\_OPEN) $\wedge$ X(response = SSH\_MSG\_CHANNEL\_OPEN\_CONFIRMATION) $\wedge$ X(request = Login) $\wedge$ X(response = SSH\_MSG\_USERAUTH\_SUCCESS)) $\rightarrow$ X(G($\neg$(response = SSH\_MSG)))) \\
\hline
11 & $SH_4$ & After the server gives a KEXINIT response, will not give the KEXINIT or AcceptConnection response until receiving the NewKeys request. & G((response = KEXINIT) $\rightarrow$ X(($\neg$(response = KEXINIT) $\wedge$ $\neg$(response = AcceptConnection)) U (request = NewKeys))) \\
\hline
12 & $SH_5$ & All authentication messages after a ConnectionSuccess response should give no
response until the end condition is true. & G((request = ConnectionSuccess) $\rightarrow$ X((NoResponse) U (EndCondition = true))) \\
\hline
13 & $SL_4$	& If SECURE\_RENEGOTIATION is disabled and the server receives a ClientHello request with renegotiation option and an empty "RENEGOTIATED\_CONNECTION" field, must send a ServerHello response without the renegotiation option. &	G((SECURE\_RENEGOTIATION = disabled) $\wedge$ (request = ClientHello) $\wedge$ (RenegotiationExtension = enabled) $\wedge$ (RENEGOTIATED\_CONNECTION = empty) $\rightarrow$ X((response = ServerHello) $\wedge$ (RenegotiationExtension = disabled))) \\
\hline
14 & $SL_5$	& If SECURE\_RENEGOTIATION is disabled the server receives a ClientHello with renegotiation extension and not an empty "RENEGOTIATED\_CONNECTION" field, must give a HandshakeFailure response. &	G((SECURE\_RENEGOTIATION = disabled) $\wedge$ (request = ClientHello) $\wedge$ (RenegotiationExtension = enabled) $\wedge$ ($\neg$(RENEGOTIATED\_CONNECTION = empty))) $\rightarrow$ X(response = HandshakeFailure))) \\
\hline
15 & $SL_6$ &	If SECURE\_RENEGOTIATION is enabled and the server receive ClientHello with SCSV (TLS\_EMPTY\_RENEGOTIATION\_INFO\_SCSV), must give a HandshakeFailure response. &	G((SECURE\_RENEGOTIATION = enabled) $\wedge$ (request = ClientHello) $\wedge$ (SCSV= enabled) $\rightarrow$ X(response = HandshakeFailure)) \\
\hline
16 & $SL_7$	& If SECURE\_RENEGOTIATION is enabled and the server receives a ClientHello request without renegotiation extension, must then abort the handshake with a HandshakeFailure response. &	G((SECURE\_RENEGOTIATION = enabled) $\wedge$ (request = ClientHello) $\wedge$ (RenegotiationExtension = false) $\rightarrow$ X(response = HandshakeFailure)) \\
\hline
17 & $SL_8$	& If SECURE\_RENEGOTIATION is enabled and the server receives ClientHello request but "RENEGOTIATED\_CONNECTION" field is not the same as the saved CLIENT\_VERIFY\_DATA value, must give a HandshakeFailure response.	& G((SECURE\_RENEGOTIATION = enabled) $\wedge$ (request = ClientHello) $\wedge$ ($\neg$(RENEGOTIATED\_CONNECTION = CLIENT\_VERIFY\_DATA))) $\rightarrow$ X(response = HandshakeFailure)) \\
\hline
18 & $SL_9$ &  {After the server receives a ClientHello request without renegotiation extension and gives a ServerHello response, then receives a ClientHello again, must refuse the renegotiation with an Alert.} &  {G(((((request = ClientHello) $\wedge$ (RenegotiationExtension = disabled)) $\wedge$ (X(response = ServerHello))) $\wedge$ (X(request = ClientHello))) $\rightarrow$ X(response = Alert))} \\
\hline
19 & $TD_{1}$ &  {If the server is in the WAIT\_CLIENTHELLO state and receives a ClientHello request with valid cookie and the epoch value 0, must finally give ServerHello responses.} &  {G(((state = WAIT\_CLINETHELLO) $\wedge$ (request = ClientHello) $\wedge$ (ValidCookie = true) $\wedge$ (EpochValue = 0) ) $\rightarrow$ X(F(response = ServerHello)))} \\
\hline
20 & $TD_{2}$ &  {If the server is in the WAIT\_CLIENTHELLO state and receives a ClientHello request with valid cookie but not 0 epoch value, must not give ServerHello responses.} &  {G(((state = WAIT\_CLINETHELLO) $\wedge$ (request = ClientHello) $\wedge$ (ValidCookie = true) $\wedge$ ( $\neg$(EpochValue = 0))) $\rightarrow$ X($\neg$(response = ServerHello)))} \\
\hline
21 & $TD_{3}$ &   {If the server is in the WAIT\_CLIENTHELLO state and receives a ClientHello request with an invalid cookie, must reply HelloVerifyRequest.} &   {G(((state = WAIT\_CLIENTHELLO) $\wedge$ (request = ClientHello) $\wedge$ (ValidCookie = false)) $\rightarrow$ X(response = HelloVerifyRequest))} \\
\hline
22 & $TD_{4}$ &   {If the server is in the WAIT\_CLIENTHELLO state but receives a ChangeCipher request, must refuse it with an Alert.} &  G(((state = WAIT\_CLIENTHELLO) $\wedge$ (request = ChangeCipher)) $\rightarrow$ X(response = Alert)) \\
\hline
23 & $TD_{5}$ & {If the server is in the DTLS\_HT\_CERTIFICATE\_REQUEST state and receives a Certificate request, must give a  DTLS\_ALERT\_HANDSHAKE\_FAILURE response or DTLS\_ALERT\_ DECODE\_ERROR, or set Client\_Auth to be verified.}  &  G(((state = DTLS\_HT\_CERTIFICATE\_REQUEST) $\wedge$ (request = Certificate)) $\rightarrow$ X((response = DTLS\_ALERT\_DECODE\_ERROR) $\vee$ (response = DTLS\_ALERT\_HANDSHAKE\_FAILURE) $\vee$ (Client\_Auth = true))) \\
\hline
24 & $TD_{6}$ & If SECURE\_RENEGOTIATION is disabled and the server receives a ClientHello request with renegotiation option and an empty "RENEGOTIATED\_CONNECTION" field, must send a ServerHello response without the renegotiation option. &	G((SECURE\_RENEGOTIATION = disabled) $\wedge$ (request = ClientHello) $\wedge$ (RenegotiationExtension = enabled) $\wedge$ (RENEGOTIATED\_CONNECTION = empty) $\rightarrow$ X((response = ServerHello) $\wedge$ (RenegotiationExtension = disabled)))) \\
\hline
25 & $TD_{7}$ & If SECURE\_RENEGOTIATION is disabled the server receives a ClientHello with renegotiation extension and not an empty "RENEGOTIATED\_CONNECTION" field, must give a HandshakeFailure response. &	G(((SECURE\_RENEGOTIATION = disabled) $\wedge$ (request = ClientHello) $\wedge$ (RenegotiationExtension = enabled) $\wedge$ ($\neg$(RENEGOTIATED\_CONNECTION = empty))) $\rightarrow$ X(response = HandshakeFailure)) \\
\hline
26 & $TD_{8}$ & If SECURE\_RENEGOTIATION is enabled and the server receive ClientHello with SCSV (TLS\_EMPTY\_RENEGOTIATION\_INFO\_SCSV), must give a HandshakeFailure response. &	G((SECURE\_RENEGOTIATION = enabled) $\wedge$ (request = ClientHello) $\wedge$ (SCSV= enabled) $\rightarrow$ X(response = HandshakeFailure)) \\
\hline
27 & $TD_{9}$ & If SECURE\_RENEGOTIATION is enabled and the server receives a ClientHello request without renegotiation extension, must then abort the handshake with a HandshakeFailure response. &	G((SECURE\_RENEGOTIATION = enabled) $\wedge$ (request = ClientHello) $\wedge$ (RenegotiationExtension = false) $\rightarrow$ X(response = HandshakeFailure)) \\
\hline
28 & $TD_{10}$ & If SECURE\_RENEGOTIATION is enabled and the server receives ClientHello request but "RENEGOTIATED\_CONNECTION" field is not the same as the saved CLIENT\_VERIFY\_DATA value, must give a HandshakeFailure response.	& G(((SECURE\_RENEGOTIATION = enabled) $\wedge$ (request = ClientHello) $\wedge$ ($\neg$(RENEGOTIATED\_CONNECTION = CLIENT\_VERIFY\_DATA))) $\rightarrow$ X(response = HandshakeFailure)) \\
\hline
29 & $TD_{11}$ & {After the server receives a ClientHello request without renegotiation extension and gives a ServerHello response, then receives a ClientHello again, must refuse the renegotiation with an Alert.} &  {G(((((request = ClientHello) $\wedge$ (RenegotiationExtension = disabled)) $\wedge$ (X(response = ServerHello))) $\wedge$ (X(request = ClientHello))) $\rightarrow$ X(response = Alert))} \\
\hline
30 & $TD_{12}$ &  {After the server receives a ClientHello request and gives a ServerHello response, then receives a ClientKeyExchange request with a different epoch value than that of ClientHello, server must not give ChangeCipherSpec responses.} &  {G((((request = ClientHello) $\wedge$ X(response = ServerHello)) $\wedge$ (X((request = ClientKeyExchange) $\wedge$ ( $\neg$($\rm EpochValue_{cke} = EpochValue_{ch}$))))) $\rightarrow$ X($\neg$(response = ChangeCipherSpec)))} \\
\hline
31 & $TD_{13}$ &  {After the server receives a ClientHello request and gives a ServerHello response, then receives a ClientHello request with the same epoch value as that of the first one, server must not give ServerHello.} &  { G((request = ClientHello) $\wedge$ (X(response = ServerHello)) $\wedge$ (X((request = ClientHello) $\wedge$ ($\rm EpochValue_{c1} = EpochValue_{c2}$))) $\rightarrow$ (X($\neg$(response = ServerHello))))} \\
\hline
32 & $TD_{14}$ & If the server receives a ClientHello request and gives a HelloVerifyRequest response, and then receives a over-large packet even with valid cookies, the server must refuse it with an Alert.  & G(((request = ClientHello) $\wedge$ X(response = HelloVerifyRequest)) $\rightarrow$ X(G((request = OverLargePacket) $\rightarrow$  X(response = Alert)))) \\
\hline
33 & $CT_{1}$ & After WILL request is received and the corresponding option is disabled, must send DO or DONT responses. & G(((request = WILL)  $\wedge$  (option = Disabled)) $\rightarrow$ X((response = DO)  $\vee$  (response = DONT))) \\
\hline
34 & $CT_{2}$ & After DO request is received and the corresponding option is disabled, must send WILL or WONT responses. & G(((request = DO)  $\wedge$  (option = Disabled)) $\rightarrow$ X((response = WILL)  $\vee$  (response = WONT))) \\
\hline
35 & $CT_3$	 & After WILL request is received and the corresponding option is enabled, must not give responses.	& G(((request = WILL) $\wedge$ (option = Enabled))  $\rightarrow$  X($\neg$(Response))) \\
\hline
36 & $CT_4$	 & After DO request is received and the corresponding option is enabled, must not give responses. & G(((request = DO) $\wedge$ (option = Enabled))  $\rightarrow$  X($\neg$(Response))) \\
\hline
37 & $CT_5$	&  After WONT request is received and the corresponding option is enabled, must send the DONT response. & 	G(((request = WONT) $\wedge$ (option = Enabled))  $\rightarrow$  X(response = DONT)) \\
\hline
38 & $CT_6$	 & After DONT request is received and the corresponding option is enabled, must send the WONT response. &	G(((request = DONT) $\wedge$ (option = Enabled))  $\rightarrow$  X(response = WONT)) \\
\hline
39 & $CT_{7}$ & After WONT request is received and the corresponding option is disabled, must not give responses. & G(((request = WONT)  $\wedge$ (option = Disabled)) $\rightarrow$ X($\neg$(Response))) \\
\hline
40 & $CT_{8}$ & After DONT request is received and the corresponding option is disabled, must not give responses. & G(((request = DONT)  $\wedge$ (option = Disabled)) $\rightarrow$ X($\neg$(Response))) \\
\hline
41 & $CT_{9}$ &	If receive IAC in NORMAL state, next go to the SIAC state and finally go back to the NORMAL state &	G((((request = IAC) $\wedge$ (state = NORMAL))  $\rightarrow$  X(G((state = IAC))) $\rightarrow$ X(F(state = NORMAL))))) \\
\hline
42 & $CT_{10}$ & Before Disconnection, must send an Alert to disconnect with clients. & G(($\neg$(Disconnection)) $\cup$  (response = Alert)) \\
\hline
43 & $CT_{11}$ & If conduct COMMAND without AbortOutput, the response must be same as the real execution results. & G(((request = COMMAND)  $\wedge$  ($\neg$(AbortOutput))) $\rightarrow$ X(G(response = realResults))) \\
\hline
44 & $PuF_1$ & If receiving invalid username or invalid password, the server must always show the same message to the user. & G(((request = InvalidUsername) $\vee$ (request = InvalidPassword)) $\rightarrow$ X(G(sameResponse))) \\
\hline
45 & $PuF_2$ & After one client succeeds to connect with a sever, the server should finally give responses for requests from the connected client. &	G((LogIN) $\rightarrow$ (X(G((Requests) $\rightarrow$ (X(F(Responses))))))) \\
\hline
46 & $PuF_3$ & If receiving the CWD request without login, the server must not give the CommandOkay response. & G(($\neg$(state = LogIN) $\wedge$ (request = CWD)) $\rightarrow$ X(G($\neg$(response = CommandOkay)))) \\
\hline
47 & $PuF_4$ & If a client doesn't log in successfully, the server must not allow this client to copy files. &	G(($\neg$(LogIN)) $\rightarrow$ (X(G((request = CopyFiles) $\rightarrow$ X($\neg$(response = CopySuccessful)))))) \\
\hline
48 & $PuF_5$ & If user directory size is larger than the set quota when the quota mechanism is activated, must finally reply that the quota is exceeded. & $\neg$F((quota\_activated = true) $\wedge$ F((user\_dir\_size \textgreater \ user\_quota)$\wedge$ G($\neg$(msg\_quota\_exceeded= true))))\\
\hline
49 & $PuF_6$ & After a connection is constructed successfully, there should be a successful login and after that without failed login. & G((((request = ValidUserName) $\wedge$ X(request = ValidPasswd)) $\wedge$ X(response = LoginSuccess)) $\rightarrow$ X(G($\neg$(response = LoginFailed))))  \\
\hline
50 & $PuF_7$ & After the connection is lost after a long time, responses should be always timeout. & G(LostConnection $\rightarrow$ X(G(response = Timeout)))\\
\bottomrule
\end{longtable}
\end{center}

\end{document}